\documentclass[12pt,letterpaper]{JHEP3}
\usepackage{amsmath,amssymb,amsfonts,xspace}
\usepackage{epsfig}

\numberwithin{equation}{section}

\def\varpi{{\bf z}}

\def\Im{\,{\rm Im}\,}
\def\Re{\,{\rm Re}\,}

\def\({\left(}
\def\){\right)}
\def\[{\left[}
\def\]{\right]}
\def\<{\left\langle}
\def\>{\right\rangle}
\def\hf{{1\over 2}}

\renewcommand{\d}{\mathrm{d}}
\newcommand{\de}{\mathrm{d}}

\newcommand{\I}{\mathrm{i}}

\newcommand{\cL}{\mathcal{L}}

\def\vrh{\varrho}

\newcommand{\p}{\partial}

\newcommand{\half}{\frac{1}{2}}

\newcommand{\cF}{\mathcal{F}}

\newcommand{\cD}{\mathcal{D}}
\newcommand{\cS}{\mathcal{S}}
\newcommand{\cG}{\mathcal{G}}
\newcommand{\cK}{\mathcal{K}}
\newcommand{\cM}{\mathcal{M}}

\newcommand{\cN}{\mathcal{N}}

\newcommand{\cX}{\mathcal{X}}
\newcommand{\CX}{\mathcal{X}}
\newcommand{\cR}{\mathcal{R}}

\newcommand{\cJ}{\mathcal{J}}
\newcommand{\cZ}{\mathcal{Z}}
\newcommand{\cI}{\mathcal{I}}
\newcommand{\cO}{\mathcal{O}}

\newcommand{\cU}{\mathcal{U}}

\DeclareSymbolFont{AMSa}{U}{msa}{m}{n}
\DeclareSymbolFont{AMSb}{U}{msb}{m}{n}
\DeclareMathSymbol{\fieldR}{\mathalpha}{AMSb}{"52}

\newcommand{\kahler}{{K\"ahler}\xspace}

\newcommand{\nn}{\nonumber}

\newcommand{\IR}{\mathbb{R}}
\newcommand{\IC}{\mathbb{C}}
\newcommand{\IZ}{\mathbb{Z}}

\newcommand{\tzeta}{\tilde\zeta}

\newcommand{\txi}{\tilde\xi}

\newcommand{\CP}{\IC P^1}

\def\bea{\begin{eqnarray}}
\def\eea{\end{eqnarray}}
\def\be{\begin{equation}}
\def\ee{\end{equation}}
\def\ba{\begin{align}}
\def\ea{\end{align}}
\def\bse{\begin{subequations}}
\def\ese{\end{subequations}}

\def\bF{\bar F}
\def\bY{\bar Y}

\def\bW{ \bar W}

\def\ba{\bar a}

\def\bw{\bar w}

\def\bz{\bar z}
\def\bu{\bar u}
\def\bv{\bar v}

\def\ze{\zeta}

\newcommand{\CL}{{\cal{L}}}

\def\ze{\zeta}

\def\ui#1{^{[#1]}}
\def\di#1{_{[#1]}}
\def\ci#1{c^{[#1]}}
\def\cij#1{c^{[#1]}}
\def\mui#1{\mu^{[#1]}}
\def\txii#1{{\tilde\xi}^{[#1]}}
\def\ai#1{{\alpha}^{[#1]}}
\def\nui#1{\nu_{[#1]}}
\def\xii#1{\xi_{[#1]}}
\def\alpi#1{\alpha^{[#1]}}
\def\Sij#1{S^{[#1]}}

\def\Hij#1{H^{[#1]}}

\def\hHij#1{\hat H^{[#1]}}
\def\hhHij#1{\hat H^{[#1]}}
\def\Xiij#1{\Xi_\gamma^{[#1]}}
\def\Xiijg#1#2{\Xi_{\gamma_{#2}}^{[#1]}}
\def\Xigi#1{\Xi_{\gamma_{#1}}}
\def\Xipgi#1{\Xi^+_{\gamma_{#1}}}
\def\Ximgi#1{\Xi^-_{\gamma_{#1}}}

\def\Hn#1{H_{\scriptscriptstyle{\smash{(#1)}}}}
\def\Hnij#1#2{\Hij{#2}_{\scriptscriptstyle{\smash{(#1)}}}}

\newcommand{\Li}{{\rm Li}}

\def\Kkl{\cK_{\gamma}}
\def\Wkl{W_{\gamma}}

\def\bWkl{\bW_{\gamma}}
\def\Thkl{\Theta_{\gamma}}
\def\Ikl{\cI_{\gamma}}
\def\Xikl{\Xi_{\gamma}}
\def\epskl{\epsilon_{\gamma}}
\def\hnkl{n_{\gamma}}
\def\Gg{\cG_{\gamma}}
\def\Ggs{G_{\gamma}}

\def\ellg#1{\ell_{#1}}

\def\Wkli#1{W_{\gamma_{#1}}}

\def\bWkli#1{\bW_{\gamma_{#1}}}
\def\Thkli#1{\Theta_{\gamma_{#1}}}

\def\Wg#1{W_{\gamma_{#1}}}

\def\bWg#1{\bW_{\gamma_{#1}}}
\def\Thg#1{\Theta_{\gamma_{#1}}}
\def\hng#1{n_{\gamma_{#1}}}
\def\Ig#1{\cI_{\gamma_{#1}}}
\def\Igg#1{\cJ_{\gamma_{#1}}}
\def\Igam#1{\cJ_{#1}}

\def\Ggss#1{G_{#1}}

\def\Sgi#1{\Sigma_{\gamma_{#1}}}

\def\derG#1{\cS_{#1}}
\def\Dlog#1{\cD_{\gamma_{#1}}\Xi}

\def\cF{\mathcal{E}}

\def\mui#1{\mu^{[#1]}}
\def\nui#1{\nu_{[#1]}}
\def\etai#1{\eta_{[#1]}}

\def\mup{\hat \mu}

\def\etap{\hat \eta}
\def\mupi#1{\hat\mu^{[#1]}}

\def\etapi#1{\hat\eta_{[#1]}}

\def\muniI#1#2#3{\mu^{[#2]#3}_{(#1)}}

\def\etaniI#1#2#3{\eta_{(#1)[#2]}^{#3}}
\def\munI#1#2{\mu^{#2}_{(#1)}}

\def\etanI#1#2{\eta_{(#1)}^{#2}}
\def\etaz{\eta_{(0)}}

\def\XXint#1#2#3{{\setbox0=\hbox{$#1{#2#3}{\int}$}
\vcenter{\hbox{$#2#3$}}\kern-.5\wd0}}

\newcommand{\cwarrow}{\text{\Large$\curvearrowright$}}
\newcommand{\ccwarrow}{\text{\Large$\curvearrowleft$}}

\def\cij#1{c}
\def\ci#1{c}

\newcommand{\hCX}{\mathcal{X}}

\def\hHij#1{H^{[#1]}}
\def\hHijg#1#2{H^{[#1]_{#2}}}

\title{D-instantons and twistors:
some exact results}
\preprint{LPTA/09-002
}
\author{Sergei Alexandrov
\\
{\it Laboratoire de Physique Th\'eorique \&
Astroparticules, CNRS UMR 5207, \\
Universit\'e Montpellier II, 34095 Montpellier Cedex 05, France}
}
\abstract{We present some results on instanton corrections to the hypermultiplet moduli
space in Calabi-Yau compactifications of Type II string theories. Previously, using twistor methods,
only a class of D-instantons (D2-instantons wrapping A-cycles) was incorporated exactly
and the rest was treated only linearly. We go beyond the linear approximation and
give a set of holomorphic functions which, through a known procedure, capture
the effect of D-instantons at all orders. Moreover, we show that for a sector where
all instanton charges have vanishing symplectic invariant scalar product,
the hypermultiplet metric can be computed explicitly.}

\begin{document}

\section{Introduction}

Compactifications of Type II string theories on Calabi--Yau (CY) threefolds
yield $\cN=2$ supergravity coupled to vector multiplets (VM) and hypermultiplets (HM).
At the two-derivative level in the effective action, the moduli spaces
of these two sectors are decoupled and can be described independently.
The vector multiplets are well known to be described by special K\"ahler geometry,
which can be conveniently encoded into
a holomorphic prepotential. The hypermultiplet sector is much more complicated
in two aspects. First, supersymmetry restricts it to be described by the so called
quaternion-K\"ahler (QK) geometry \cite{Bagger:1983tt}. Contrary to the K\"ahler geometry,
QK spaces are not described by any potential and thus
it is difficult to parametrize them in a simple way.
Second, since the HM include the dilaton, the effective
action in the hypermultiplet sector receives corrections in the string coupling
$g_s$, both perturbative and non-perturbative. The non-perturbative corrections
are related to D-instantons arising as D-branes wrapping non-trivial cycles on
the internal CY \cite{Becker:1995kb}. But due to the lack of a well established instanton calculus
in string theory, the direct calculation of such D-instanton corrections
seems to be a very hard task. As a result, the exact metric on the HM moduli space
remains still unknown and represents a great challenge in string theory.

Recently, a large progress in this direction has been achieved.
A crucial step was provided by understanding of how to overcome
the first of the above mentioned problems. Namely, it was realized
that a suitable parametrization of QK spaces can be found using the so called
twistor techniques \cite{MR664330,MR1096180}, which in the physics literature
appeared as the projective superspace method
\cite{Hitchin:1986ea,Lindstrom:1987ks,Ivanov:1995cy,deWit:2001dj,Lindstrom:2008gs}.

In particular, it was known for long time that
$4d$-dimensional QK manifolds $\cM$ are in one-to-one correspondence with
$4d+4$-dimensional hyperk\"ahler (HK) cones, or Swann bundles $\cS$ \cite{MR1096180},
which are hyperk\"ahler manifolds with an additional homothetic Killing vector
and an isometric $SU(2)$ action.
The geometry of a HK manifold in turn can be encoded
into the complex symplectic structure
on its twistor space $\cZ_\cS$. It turns out that the latter is characterized by a set of
holomorphic functions, which generate symplectomorphisms between local Darboux coordinates
in various patches covering the twistor space \cite{Alexandrov:2008ds}.
These functions play the same role for HK geometry as the holomorphic prepotential for the
special K\"ahler geometry. However, the additional structure of HK cone imposes
some restrictions on these transition functions, which allow to descend to the twistor space
$\cZ$ of the initial QK manifold \cite{Alexandrov:2008nk}.
As a result, the QK geometry can be obtained directly from the
complex contact structure on $\cZ$ \cite{lebrun1989m} and is characterized by a related set of
holomorphic functions interpreted now as complex contact transformations between
different locally flat patches.

On the other hand, the absence of string instanton calculus was overcome by
applying various non-perturbative (self)dualities of Type IIA and Type IIB string theories.
In this way, applying $SL(2,\IZ)$ self-duality of Type IIB string theory to the perturbative HM metric
found before in \cite{Rocek:2005ij,Robles-Llana:2006ez,Alexandrov:2007ec}
(see also \cite{Berkovits:1995cb,Berkovits:1998jh,Antoniadis:1997eg,Gunther:1998sc,Antoniadis:2003sw}
for some earlier work),
the authors of \cite{RoblesLlana:2006is} were able to
compute the D(-1) and D1-instanton corrections. Mirror symmetry further maps these corrections
into D2-brane instantons of Type IIA theory compactified on the mirror CY threefold $X$ \cite{RoblesLlana:2007ae}.
However, this map recovers only instantons wrapping A-cycles in some symplectic polarization
of $H_3(X)$. The contributions of other D2 instantons may then be
restored using the symplectic invariance  of Type IIA theory \cite{Alexandrov:2008gh}.

For these developments, the twistor and projective techniques mentioned above were indispensable
since they allowed to work with simple holomorphic functions encoding in a concise way
complicated QK metrics.
There is however an important difference in the twistor description of A-type D2 instantons and
more general ones.
Although the transition functions on the twistor space $\cZ$ define the metric on
the associated QK space $\cM$ in a unique way, this requires the knowledge of the so called
contact twistor lines, representing local Darboux coordinates on $\cZ$ in terms of
coordinates on $\cM$ and the coordinate $\varpi$ on the $\CP$ fiber.
Generically, this requires solution of complicated integral equations and cannot
be done explicitly.
But in the special case when the QK space has $d+1$ commuting isometries, the solution
is known and can be described by the so called $\cO(2)$ multiplets. This is precisely the case
of the HM moduli space with only A-type D2 instantons included.
All other instantons break the isometries to discrete subgroups and
a solution for the twistor lines can be written only as a perturbative series in instantons.
In particular, the linear approximation was explicitly found in \cite{Alexandrov:2008gh}.

The aim of this paper is to present some {\it exact} results, going beyond the linear approximation.
The paper consists from two parts.
The first part concerns the underlying mathematical construction, whereas the second
addresses the physical problem of the HM moduli space.

The first part gives mostly a very brief review of the previous works
\cite{Alexandrov:2008ds,Alexandrov:2008nk}, introducing the necessary notions and notations.
It is split into two parts dealing with HK and QK cases, respectively.
Besides the simple review, it contains also some new results.
In particular, we derive the formula \eqref{defL} for the exact
``Lagrangian", related to the K\"ahler potential on HK space by
Legendre transform, and establish a relation with the projected superspace
formulation of \cite{Lindstrom:2008gs}.

In the second part we elaborate on the results of \cite{Alexandrov:2008gh},
where the contributions of all D2-instantons were found in the linear approximation.
First, we show that with a suitable choice of transition functions, favored
by symplectic invariance, in the case of a single D2-instanton the linear approximation
becomes exact. It is not something unexpected because by a suitable symplectic transformation
any D2-instanton can be mapped to a D2-brane wrapping an A-cycle, for which the description
in terms of $\cO(2)$ multiplets is applied
and the contact twistor lines are known exactly. Thus, our result can be considered as
a non-trivial test satisfied by the proposed transition functions.

Then the inclusion of several charges is considered
and, although we are not able to compute explicitly the twistor lines,
we present some new insights. In particular, we give a full consistent set of transition
functions incorporating all D-instantons, improving
a somewhat naive proposal of \cite{Alexandrov:2008gh}.
This set is necessary for the complete description
of the HM moduli space and its twistor space.
Furthermore, we find that a certain sector, consisting from D2-instantons
with all ``mutually local charges", admits the exact description similar to the one of the single-charge case.
Finally, we discuss the consistency of our results with symplectic invariance
and the so called wall-crossing conditions \cite{ks,Gaiotto:2008cd}.

\section{Twistor description of HK and QK spaces}

\subsection{K\"ahler potential for HK manifolds}

We start from the twistor approach to description of HK spaces.
Although it is not directly needed for our discussion of the HM moduli space in
the following sections, we include it to present an explicit formula for the K\"ahler potential
in terms of twistor lines. Our discussion closely follows \cite{Alexandrov:2008ds}.

The twistor space is a $\CP$ bundle over the initial HK manifold.
It is equipped with a complex structure and a holomorphic two-form
\be
\Omega(\zeta) = \omega^+ -\I \zeta \omega^3 + \zeta^2 \omega^-,
\label{Omeg}
\ee
where $\zeta$ is a coordinate on $\CP$ and $\omega^i$ are K\"ahler forms
associated with three complex structures $J^i$ carried by any HK manifold.
Note that $\omega^+$ ($\omega^-$) is (anti-)holomorphic with respect to the complex structure $J^3$.

In fact, $\Omega$ is a section of a two-form valued $\cO(2)$ bundle on $\CP$ \cite{Hitchin:1986ea}
which is reflected in the fact that the representation \eqref{Omeg} diverges at $\zeta=\infty$.
This signifies that one should cover $\CP$ by two open intersecting patches, $\cU_0$ and $\cU_\infty$,
around the north and south poles, respectively, and the holomorphic two-form is represented
in every patch as
\be
\Omega\ui{0}(\zeta)=\Omega(\zeta),
\qquad
\Omega\ui{\infty}(\zeta)= f_{0\infty}^{-2}\, \Omega(\zeta)
= \omega^- -\I\ze^{-1} \omega^3+ \ze^{-2} \omega^+ ,
\label{Ominfty}
\ee
where $f_{0\infty}^2=\zeta^2$ is the transition function of the line bundle $\cO(2)$ over $\CP$.
More generally, we have to work with a set of patches $\cU_i$ and representatives $\Omega^{[i]}$
such that on every overlap $\cU_i \cap \cU_j$ they are related by
\be
\label{omij}
\Omega^{[i]}= f_{ij} ^2 \, \Omega^{[j]}  \,\quad \mod\, \de\zeta .
\ee

In every patch one can introduce a local system of Darboux coordinates
\be
\label{darboux}
\Omega^{[i]}=\de\mui{i}_I\wedge \de \nui{i}^I .
\ee
Different systems on the overlaps of two patches are related by symplectomorphisms,
which can be expressed through generating functions $\Sij{ij}(\nui{i},\mui{j},\zeta)$
of the initial ``position" and final ``momentum" coordinates.
Then the gluing conditions take the following form
\be
\nui{j}^I = \p_{\mui{j}_I}\Sij{ij}(\nui{i},\mui{j},\zeta) , \qquad
\mui{i}_I =f_{ij}^2\,\p_{\nui{i}^I}\Sij{ij}(\nui{i},\mui{j},\zeta)  .
\label{cantr}
\ee
The transition functions $\Sij{ij}$ contain all information about the
twistor space and the underlying HK manifold. They can be chosen arbitrarily
up to some consistency and reality conditions, and define the HK space
uniquely up to some gauge freedom \cite{Alexandrov:2008ds}. All geometric
information can be restored once we solved the gluing conditions \eqref{cantr}
for the coordinates $\nu^I$ and $\mu_I$ as functions of $\zeta$. Such functions are
called twistor lines. The free parameters of the solution play the role
of coordinates on the HK base.

Here we will consider the situation where $\nu^I$ and $\mu_I$
are perturbations of global sections of $\cO(2)$ and $\cO(0)$, respectively.
Then it is convenient to redefine
\be
\etai{i}^I(\zeta) \equiv \zeta^{-1}f_{0i}^2\nui{i}^I(\zeta)
\label{defeta}
\ee
and choose the generating functions as
\be
\Sij{ij}(\nui{i},\mui{j},\zeta)=\zeta f_{0j}^{-2}\(\etai{i}^I\mui{j}_I
-\Hij{ij}(\etai{i},\mui{j},\zeta) \) .
\label{genf}
\ee
However, $\Hij{ij}$ are not assumed to be infinitesimal and therefore
we do not actually impose any restrictions.
The gluing conditions then become
\be
\label{smalleq}
\etai{j}^I=\etai{i}^I-\p_{\mui{j}_I}\Hij{ij},
\qquad
\mui{j}_I=\mui{i}_I+\p_{\etai{i}^I}\Hij{ij} .
\ee
These conditions can be rewritten as integral equations, which are suitable for
the perturbative treatment (see appendix \ref{ap_pertsol})
\be
\begin{split}
\etai{i}^I =&\, \etaz^I +\etapi{i}^I,
\qquad
\etapi{i}^I=-\hf \sum_j\oint_{C_j} \frac{\de\zeta'}{2\pi \I\,\zeta'}\,
\frac{\zeta'^3+\zeta^3}{\zeta\zeta' (\zeta'-\zeta)}\, H^{[ij]I}(\zeta') ,
\\
\mui{i}_I =&\,\frac{\I}{2}\,\vrh_I+\mupi{i}_I,
\qquad
\mupi{i}_I=\hf\sum_j\oint_{C_j} \frac{\de\zeta'}{2\pi \I\,\zeta'}\,
\frac{\zeta'+\zeta}{\zeta'-\zeta}\, \Hij{ij}_I(\zeta') ,
\end{split}
\label{soltwist}
\ee
where the variable $\zeta$ is inside the contour $C_i$ surrounding $\cU_i$ in
the counterclockwise direction,
$H^{[ij]I}\equiv \p_{\mui{j}_I}\Hij{ij}$, $\Hij{ij}_I\equiv\p_{\etai{i}^I}\Hij{ij}$ and
\be
\etaz^I(\zeta) = \frac{v^I}{\zeta} + x^I - \bv^I \zeta .
\label{omult}
\ee

Of course, the integral equations \eqref{soltwist} are as difficult as the initial equations
\eqref{smalleq}. But on the other hand, they introduce explicitly the coordinates on the HK base,
$v^I, \bv^I, x^I$ and $\vrh_I$, and
allow to get the complex structure together with the K\"ahler potential.
Indeed, due to \eqref{Omeg}, the holomorphic form $\omega^+$ is obtained as the constant term in the small
$\zeta$-expansion of \eqref{darboux} in the patch $\cU_0$
\be
\label{omplus}
\omega^+
=\de\(\frac{\I}{2}\, \vrh_I+\hf\oint_C \frac{\de\zeta}{2\pi \I\, \zeta}\, H_I \)
\wedge \de\(v^I-\hf \oint_C \frac{\de\zeta}{2\pi \I}\, H^I \)   ,
\ee
where, to avoid cluttering, we omitted the sum over different patches.
From this result, we read the complex coordinates on our HK space
\be
u^I \equiv   v^I-\hf \oint_C \frac{\de\zeta}{2\pi \I}\, H^I  ,
\qquad
w_I \equiv  \frac{\I}{2}\, \vrh_I+\hf\oint_C \frac{\de\zeta}{2\pi \I\, \zeta}\, H_I .
\label{complcoor}
\ee

Similarly one can get a formula for the K\"ahler form $\omega^3$. It imposes the following conditions
on the K\"ahler potential (the index on $K$ denotes the derivative w.r.t.
the corresponding variable)
\be
\label{dK}
K_{u^I} = \oint_C \frac{\de\zeta}{2\pi \I\, \zeta^2}\, H_I  ,
\qquad
K_{w_I} = - x^I +\hf\oint_C \frac{\de\zeta}{2\pi \I\,\zeta}\, H^I .
\ee
Let us trade the K\"ahler potential for its Legendre transform
\be
\label{Kdef}
K(u,\bu,w,\bw)= \langle \CL(u,\bu,x,\vrh) -x^I (w_I+\bw_I) \rangle_{x^I} .
\ee
In the projective approach the function $\CL$ appears as a Lagrangian on
the projective superspace \cite{Hitchin:1986ea,Ivanov:1995cy,Lindstrom:2008gs}.
From \eqref{dK}, the Lagrangian must satisfy
\be
\cL_{u^I} = \oint_C \frac{\de\zeta}{2\pi \I\, \zeta^2}\, H_I  ,
\qquad
\cL_{x^I} = \oint_C \frac{\de\zeta}{2\pi \I\, \zeta}\, H_I  ,
\qquad
\cL_{\vrh_I} = \frac{\I}{2}\oint_C \frac{\de\zeta}{2\pi \I\,\zeta}\, H^I .
\label{condLang}
\ee
Remarkably, these equations are integrable (see appendix \ref{ap_Ksol})
and solved by
\be
\cL=\oint_C \frac{\de\zeta}{2\pi \I\,\zeta}\(H-\mup_I\p_{\mu_I}H\).
\label{defL}
\ee
Together with \eqref{Kdef}, this result gives an explicit representation for the K\"ahler potential.

Our result can also be rewritten in the form found in \cite{Lindstrom:2008gs}.
Using \eqref{smalleq} and \eqref{complcoor}, it is easy to check that
\be
K=-\oint_C \frac{\de\zeta}{2\pi \I\,\zeta}\, \tilde S
-w_I \oint_{C_0} \frac{\de\zeta}{2\pi \I\,\zeta}\,\etai{0}^I
+\bw_I \oint_{C_\infty} \frac{\de\zeta}{2\pi \I\,\zeta}\,\etai{\infty}^I,
\ee
where
\be
\tilde S^{[ij]}(\etai{i},\etai{j})=
\zeta^{-1} f_{0j}^2\<\Sij{ij}(\nui{i},\mui{j},\zeta)-\nui{j}^I\mui{j}_I\>_{\mui{j}_I}
\ee
are generating functions of symplectomorphisms written as functions of two ``positions".
This representation coincides with eq. (4.1) of \cite{Lindstrom:2008gs}, where the case
of only two patches was considered, provided $\zeta\etai{0}^I$ and $-\zeta^{-1}\etai{\infty}^I$
are identified as arctic and antarctic multiplets, respectively.

As usual, the metric can be computed without
knowing $x^I$ as a function of $u^I, \bu^I,w_I, \bw_I$ explicitly. It is expressed through
the derivatives of the Lagrangian as
\be\label{HKmet}
\begin{split}
& K_{u^I \bu^J} = \CL_{u^I \bu^J}-\CL_{u^I x^K}\CL^{x^K x^L}\CL_{x^L \bu^J} ,
\\
& K_{u^I \bw_J} =\CL_{u^I x^K}\CL^{x^K x^J}+\I\[\CL_{u^I\vrh_J}-\CL_{u^I x^K}\CL^{x^K x^L}\CL_{x^L \vrh_J}\] ,
\\
& K_{w_I \bu^J} = \CL^{x^I x^K}\CL_{x^K \bu^J}-\I\[\CL_{\vrh_I \bu^J}-\CL_{\vrh_I x^K}\CL^{x^K x^L}\CL_{x^L \bu^J}\]  ,
\\
& K_{w_I \bw_J} = \CL_{\vrh_I\vrh_J}-\cL_{\vrh_I x^K}\cL^{x^K x^L} \cL_{x^L\vrh_J}-\CL^{x^I x^J}
+\I\[\CL^{x^I x^K}\CL_{x^K \vrh_J}-\CL_{\vrh_I x^K}\CL^{x^K x^J}\]  ,
\end{split}
\ee
where $\CL^{x^Ix^J}$ denotes the inverse of the matrix $\CL_{x^Ix^J}$.
In Appendix  \ref{ap_Ksol} we provide also
formulae for the inverse metric.

\subsection{Twistor space of QK spaces}

As was mentioned in the Introduction, QK spaces are in one-to-one correspondence
with HK cones and thus can be described by the formalism of the previous subsection,
provided one imposes additional constraints taking into account that the HK space
in question is a cone. In particular, the transition functions $\Hij{ij}$ must be
homogeneous of first degree in the arguments $\etai{i}^I$ and independent of
$\zeta$ \cite{Kuzenko:2007qy,Alexandrov:2008nk,Ionas:2008gh}.
On the other hand, in \cite{lebrun1989m,Alexandrov:2008nk}
it was shown that QK spaces can be described directly in terms of
its twistor spaces $\cZ$ and all geometric information is again contained in
transition functions relating different sets of coordinates analogous to the
Darboux coordinates from the previous subsection.
In this case, these are canonical coordinates for the contact one-form
\be
\CX\ui{i}  \equiv  \de \alpi{i}
+  \xii{i}^\Lambda \, \de \txii{i}_\Lambda.
\label{con1fo}
\ee
They can be very easily related to the coordinates $\nu^I, \mu_I$ used above and
the $\cO(2)$-valued complex Liouville form on $\cZ_\cS$:
\be
\label{contact1}
\xii{i}^\Lambda = {\nui{i}^\Lambda}/{\nui{i}^\alpha},
\qquad
\txii{i}_\Lambda = \mui{i}_\Lambda,
\qquad
\alpi{i}= \mui{i}_\alpha,
\qquad
\nui{i}^\alpha\CX\ui{i}={\nu^I_{[i]} {\rm d} \mu_I^{[i]}}{} ,
\ee
where we have singled out one coordinate
$\nui{i}^\alpha$, and denoted by $\nui{i}^\Lambda$ the remaining $d$ coordinates.
One can show that the HK cone conditions ensure that
$\xii{i}^\Lambda, \txii{i}_\Lambda$ and $\ai{i}$ are all functions of the coordinates $x^\mu$ on $\cM$
and the coordinate $\varpi$ parametrizing the $\CP$ fiber of $\cZ$.

On the overlap of two patches, the contact form satisfies \cite{Alexandrov:2008nk}
\be
\label{glue2}
\CX\ui{i} =  \hat f_{ij}^{2} \, \CX\ui{j} ,
\qquad
\hat f_{ij}^{2}\equiv f_{ij}^{2} \, \nui{j}^\flat / \nui{i}^\flat =\etai{j}^\flat/\etai{i}^\flat,
\ee
whereas the canonical coordinates are related
by contact transformations
\be
\begin{split}
\xii{j}^\Lambda = &  \xii{i}^\Lambda + T_{[ij]}^\Lambda,
\qquad
\txii{j}_\Lambda =  \txii{i}_\Lambda
 + \tilde{T}^{[ij]}_\Lambda ,
\qquad
\ai{j} =  \ai{i}
 + \tilde{T}^{[ij]}_\alpha ,
\end{split}
\label{QKgluing}
\ee
where we abbreviated
\be
\begin{split}
\label{Tfct}
& T_{[ij]}^\Lambda \equiv
 -\p_{\txii{j}_\Lambda }\hHij{ij}
+\xii{j}^\Lambda \, \p_{\ai{j} }\hHij{ij}  ,
\\
\tilde{T}^{[ij]}_\Lambda \equiv &
 \p_{\xii{i}^\Lambda } \hHij{ij} ,
\qquad
\tilde{T}^{[ij]}_\alpha\equiv
\hHij{ij}- \xii{i}^\Lambda \p_{\xii{i}^\Lambda}\hHij{ij}
\end{split}
\ee
with $\hHij{ij}$ being a general function of $\xii{i}^\Lambda, \txii{j}_\Lambda$ and $\ai{j}$.
Moreover, the coefficients $\hat f_{ij}^2$, appearing in \eqref{glue2},
are also determined by $\hHij{ij}$,
\be
\hat f_{ij}^2=1-\p_{\ai{j} }\hHij{ij}.
\label{trans_f}
\ee

To deal with the gluing conditions, we assume that the section $\nu^\flat$ on the twistor space
$\cZ_\cS$ of the Swann bundle has only two zeros
situated at the centers of patches $\cU_\pm$.
(This assumption is valid in the important particular case when $\nu^\flat$ is a global $\cO(2)$ section.)
By an SU(2) rotation one can always bring them to the points
$\varpi=0$ and $\varpi=\infty$, respectively. Then these points correspond to simple poles
of $\xii{\pm}^\Lambda$. These are the only singularities of the contact Darboux coordinates
$\xii{i}^\Lambda, \txii{i}_\Lambda,\ai{i}$ except the ones introduced by
the so called ``anomalous dimensions" $c_\Lambda,c_\alpha$. The latter originate
from possible logarithmic branch cuts on $\cZ_\cS$ and represent additional input
supplementing transition functions \cite{Alexandrov:2008nk}.
As a result, the gluing conditions \eqref{QKgluing} for the contact Darboux coordinates
can be written as the following integral equations
\bea
\xii{i}^\Lambda(\varpi,x^\mu)& =& A^\Lambda +
\varpi^{-1} Y^\Lambda - \varpi \bY^\Lambda
+\frac12 \sum_j \oint_{C_j}\frac{\de\varpi'}{2\pi\I \varpi'}\,
\frac{\varpi'+\varpi}{\varpi'-\varpi}\, T_{[+j]}^\Lambda(\varpi') ,
\nonumber \\
\txi_\Lambda^{[i]}(\varpi,x^\mu)& = &\frac{\I}{2}\, B_\Lambda +
\half  \sum_j \oint_{C_j} \frac{\de \varpi'}{2 \pi \I \varpi'} \,
\frac{\varpi' + \varpi}{\varpi' - \varpi}
\, \tilde{T}_\Lambda^{[+j]}(\varpi')+  \ci{+}_\Lambda \log \varpi ,
\label{txiqline}
\\
\ai{i}(\varpi,x^\mu)& = &\frac{\I}{2}\, B_\alpha +
\half  \sum_j \oint_{C_j} \frac{\de \varpi'}{2 \pi \I \varpi'} \,
\frac{\varpi' + \varpi}{\varpi' - \varpi}
\, \tilde{T}^{[+j]}_\alpha(\varpi')+  \ci{+}_\alpha  \log \varpi
+\cij{+}_\Lambda\(Y^\Lambda \varpi^{-1} + \bY^\Lambda \varpi\),
\nonumber
\eea
where $Y^\Lambda, A^\Lambda, B_\Lambda, B_\alpha$ are free parameters playing the role
of coordinates on $\cM$. They contain one parameter more than the dimension of $\cM$
because the overall phase rotation of $Y^\Lambda$ can be absorbed by a redefinition
of the $\CP$ coordinate $\varpi$.

Similarly to the holomorphic two-form on $\cZ_\cS$ \eqref{Omeg}, the contact form on $\cZ$
is restricted to have the following expansion
\be
\label{contact}
\hCX\ui{i} = 2\,  \frac{e^{\Phi\di{i}}}{\varpi}
\(\de\varpi + p_+ -\I p_3 \,\varpi + p_-\, \varpi^2\) ,
\ee
where $\vec p$ is the $SU(2)$ part of the Levi-Civita connection on $\cM$
and the function $\Phi\di{i}$ is the so called ``contact potential".
In general it is defined only locally and therefore carries the index of the patch $\cU_i$.
This is an important object since its real part provides a \kahler potential
for the \kahler-Einstein metric on $\cZ$
\be
\label{Knuflat}
K_{\cZ}\ui{i} = \log\frac{1+\varpi\bar \varpi}{|\varpi|}
+ \Re\Phi\di{i}(x^\mu,\varpi) .
\ee
The gluing conditions for the contact potential are determined by \eqref{trans_f},
\be
\Phi_{[i]}-\Phi_{[j]}  =  \log\(1- \p_{\ai{j} }\hHij{ij}\)
\label{gluPhi}
\ee
and the potential must be regular everywhere. The condition \eqref{gluPhi}
can be solved in terms of solutions of \eqref{txiqline} as 
\be
\Phi_{[i]}=\phi-\frac12 \sum_j \oint_{C_j} \frac{\de \varpi'}{2 \pi \I \varpi'}
\,\frac{\varpi' + \varpi}{\varpi' - \varpi}\, \log\(1- \p_{\ai{j} }\hHij{+j}(\varpi')\).
\label{solcontpot}
\ee
The first term in \eqref{solcontpot}, which is a real constant,
can actually be computed explicitly using \eqref{contact} and expanding the Darboux coordinates \eqref{txiqline}
near $\varpi=0$.  The result can be given in one of the two forms\footnote{To get the second form from the first, 
one should use the relation
$$
\sum_j\oint_{C_j}\frac{\de\varpi}{\varpi}\[
\(\varpi^{-1} Y^{\Lambda}+\varpi \bY^{\Lambda} \)\tilde{T}_\Lambda^{[+j]}
+\cij{+}_\Lambda T_{[+j]}^\Lambda
-2 e^{\Phi_{[j]}}\p_{\ai{j}}\hHij{+j}\]=0,
$$
which holds since the l.h.s. is just an integral of a total derivative, 
and the following remarkable identity
$$
\frac{1}{4\pi}\sum_j\oint_{C_j}
\frac{\de\varpi}{\varpi}
\p_{\ai{j}}\hHij{+j} e^{-\frac12 \sum_k \oint_{C_k} \frac{\de \varpi'}{2 \pi \I \varpi'}
\,\frac{\varpi' + \varpi}{\varpi' - \varpi}\, \log\(1- \p_{\ai{k} }\hHij{+k}(\varpi')\)} 
=-\sin\[ \frac{1}{4\pi} \sum_j\oint_{C_j}\frac{\de\varpi}{\varpi}\,\log\(1-\p_{\ai{j}}\hHij{+j} \)\].
$$
}
\be
\begin{split}
e^\phi=&\, 
\hf\left|\frac12\sum_j\oint_{C_j}\frac{\de\varpi}{2\pi\I \varpi^2}\, Y^\Lambda \p_{\xii{+}^\Lambda}\hHij{+j}
+{\cij{+}_\Lambda}\( A^\Lambda+\hf\sum_j \oint_{C_j}\frac{\de\varpi}{2\pi\I \varpi}\,
T_{[+j]}^\Lambda\)
+{\cij{+}_\alpha}\right|
\\
=&\,\displaystyle{
\frac{\frac12 \sum_j\oint_{C_j}\frac{\de\varpi}{2\pi\I \varpi}
\(\varpi^{-1} Y^{\Lambda}-\varpi \bY^{\Lambda} \)
\p_{\xii{+}^\Lambda}\hHij{+j}
+ \cij{+}_\Lambda A^\Lambda + \cij{+}_\alpha}
{2\cos\[ \frac{1}{4\pi} \sum_j\oint_{C_j}\frac{\de\varpi}{\varpi}\,\log\(1-\p_{\ai{j}}\hHij{+j} \)\]}.
}
\end{split}
\label{contpotconst}
\ee
Note that the denominator in the second representation is quadratic in perturbations
around the case with $d+1$ commuting isometries (the non-perturbed transition functions
$\hHij{ij}$ are independent of $\txi_\Lambda$ and $\alpha$) so that our result agrees with the one 
obtained in \cite{Alexandrov:2008nk} in the linear approximation.

The contact twistor lines and the contact potential \eqref{solcontpot}
provide sufficient information to compute the metric on $\cM$.
A procedure to do this is described in detail in \cite{Alexandrov:2008nk,Alexandrov:2008gh}.
Note that in the important particular case where the transition functions
are independent of $\alpha$, the above description crucially simplifies:
$\hat f_{ij}^2=1$ and the contact potential is globally defined
coinciding with its constant part $\phi$ which also acquires significant simplifications.
This is the case for the hypermultiplet moduli space in the absence of NS5-brane
instantons considered in the next section.

\section{D-instantons and hypermultiplet moduli space}

\subsection{HM moduli space in Type IIA string theory}

Our aim is to investigate the HM moduli space of Type IIA string theory compactified on
a CY threefold. It comprises $h_{2,1}(X)+1$ hypermultiplets, which include
the complex structure moduli
$X^\Lambda=\int_{\gamma^\Lambda} \Omega$,
$F_\Lambda=\int_{\gamma_\Lambda} \Omega$,
the RR scalars $\zeta^\Lambda, \tzeta_\Lambda$ representing the RR three-form
integrated along a symplectic basis $(\gamma^\Lambda,\gamma_\Lambda)$
of A and B cycles in $H_3(X,\IZ)$, the four-dimensional dilaton
$e^{\phi}=1/g_{(4)}^2$ and the Neveu-Schwarz (NS) axion $\sigma$, dual to the
Neveu-Schwarz two-form $B$ in four dimensions. Whereas $X^\Lambda$ provide a set
of homogeneous coordinates for complex structure deformations,
they may be traded for the inhomogeneous coordinates $z^a=X^a/X^0$.

To describe the HM moduli space, we will use the twistor approach from the previous section.
This means that we should provide a covering of $\CP$, an associated set
of transition functions and a set of anomalous dimensions.
These data allow in principle
to compute the contact twistor lines \eqref{txiqline},
the contact potential \eqref{solcontpot} and to derive the metric.
The coordinates appearing as free parameters of these solutions can be related
to the physical fields of Type IIA string theory using considerations of symplectic invariance
\cite{Alexandrov:2008gh}.

For the perturbative metric, which was initially obtained
via the projective superspace and superconformal quotient
\cite{Rocek:2005ij,Robles-Llana:2006ez,Neitzke:2007ke,Alexandrov:2007ec},
such a formulation was given in  \cite{Alexandrov:2008nk}.
At tree level, the HM moduli space is determined via the ``c-map"
construction \cite{Cecotti:1989qn,Ferrara:1989ik} from the moduli space
of complex structure deformations and therefore it is completely characterized
by the holomorphic prepotential $F(X)$.
It specifies the transition functions $\hHij{\pm 0}$ corresponding to the
covering of the twistor space $\cZ$ by three patches: two patches $\cU_+$, $\cU_-$
are open disks centered around $\varpi=0$ and $\varpi=\infty$
and a third patch $\cU_0$ covers the rest of $\CP$ (see the next subsection for
precise formulae).
The one-loop correction is determined by the Euler number of CY,
$\chi_X=2(h^{1,1}(X)-h^{2,1}(X))$, and incorporated through
a non-vanishing anomalous dimension $\cij{+}_\alpha$.
It has been argued that there are no higher loop corrections \cite{Robles-Llana:2006ez}.

The inclusion of non-perturbative effects into this framework was considered in
\cite{Alexandrov:2008gh}. It was suggested that every D2-instanton of charge
$\gamma=(q_\Lambda,p^\Lambda)$, {\it i.e.} wrapping a three-cycle in the homology class
$q_\Lambda\gamma^\Lambda-p^\Lambda\gamma_\Lambda\in H_3(X,\IZ)$,
defines two ``BPS rays" on $\CP$ going from $\varpi=0$ to $\varpi=\infty$.
These rays introduce discontinuities in the Darboux coordinates, thus requiring introduction
of additional patches.
The transition functions through the rays are determined
by dilogarithm functions, $\Li_2(x)=\sum_{m=1}^\infty m^{-2}x^m$, of symplectic invariant
combinations of Darboux coordinates, whereas the weights of instanton contributions were
argued to be given by generalized Donaldson-Thomas invariants found
in \cite{ks}.\footnote{A similar picture was suggested also in \cite{Gaiotto:2008cd}
in the context of $D=3,\cN=4$ supersymmetric gauge theories obtained by compactifying
$D=4,\cN=2$ theories on a circle. In this case,
the moduli space is HK and corrected by instantons representing
4D BPS solitons winding around the compactification circle.
In fact, the two moduli spaces are very close not only at the qualitative level, but also
quantitatively since the D-instantons are described in both cases by essentially the same
transition functions.}

However, these instanton corrections were analyzed only in the linear
approximation. Although a proposal for the transition functions describing
the exact twistor space has also been given, we will see that it was not quite precise and complete.
In particular, only transition functions through the BPS rays were proposed,
but even they were oversimplified.
Below we present such an exact complete description providing
a full consistent set of transition functions, which are argued to be valid
at all orders in the instanton expansion. Our argumentation is based
on symplectic invariance and it will be discussed in detail in the end of this section
when all essential properties of the construction are already deduced.
The construction itself represents the core of this section.
First, we show how it is realized in the case of a single D2-instanton, where
all relevant quantities can be computed explicitly, and then we generalize it
to incorporate all instantons.
Moreover we find that for a subset of all D-instantons consisting of ``mutually local states"
there is an explicit representation of the contact twistor lines similar to the single-charge
case.

\subsection{The case of a single charge}
\label{subsec_single}

\subsubsection{Twistor space in the presence of one D2-instanton}

In this subsection we consider the situation when the HM moduli space
is affected by only one D2-instanton of charge $\gamma=(q_\Lambda,p^\Lambda)$.
We give an improved version of the twistor space suggested in
\cite{Alexandrov:2008gh} and compute the exact contact twistor lines and the contact potential.

\EPSFIGURE{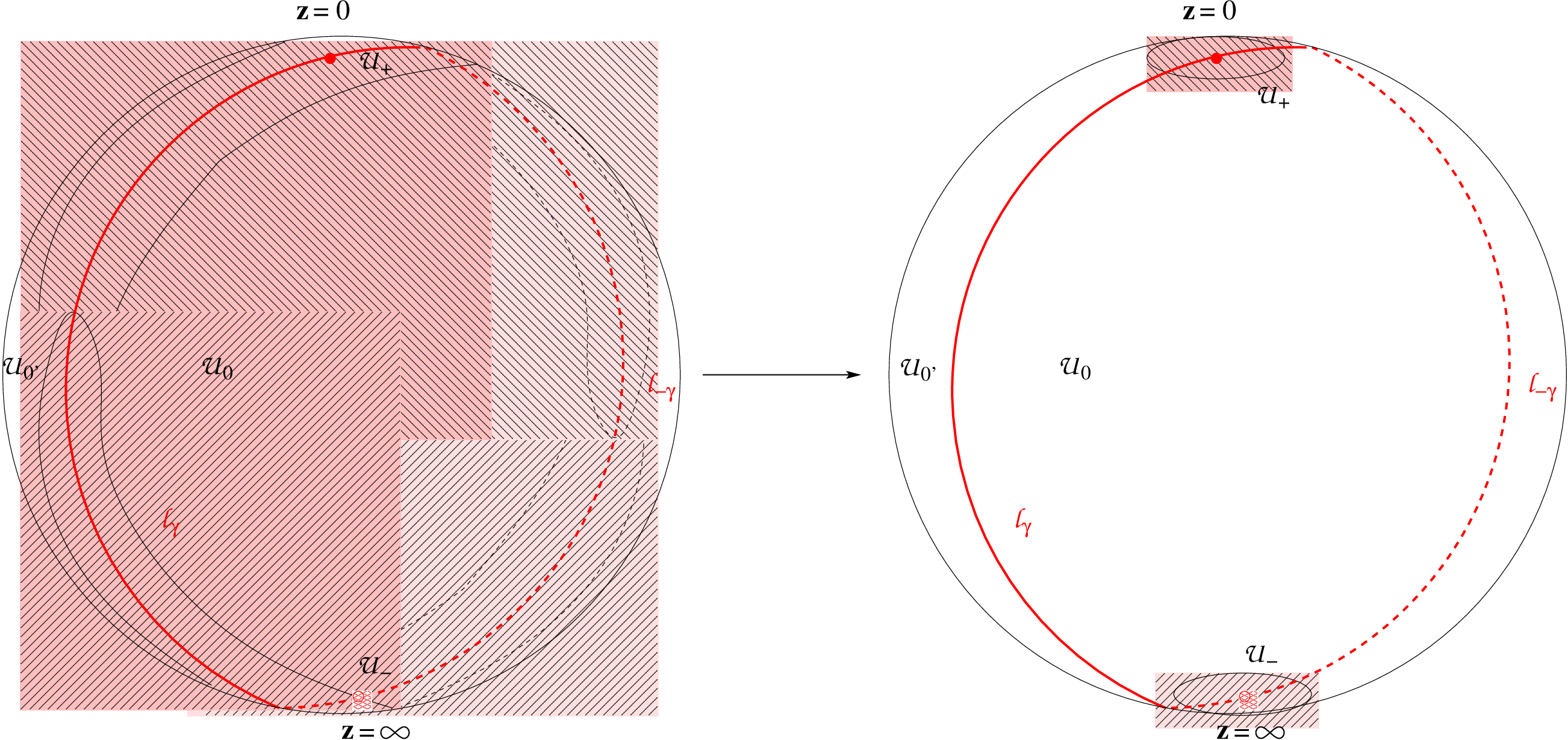,height=7.5cm}{Two coverings of $\CP$. The
covering on the left is at the basis of our construction of the HM twistor space
affected by one D-instanton. The covering on the right
is obtained in the limit where the strips $\cU_\pm$ go to
zero width along the meridians $\ellg{\pm\gamma}$, while maintaining a non-zero size at
the north and south pole. \label{fig_sphere}}

First we introduce the following covering of $\CP$ (see Fig. \ref{fig_sphere}, left).
As we mentioned above, each charge vector
$\gamma$ defines a pair of ``BPS rays''
$\ellg{\pm \gamma}$ on $\CP$ going between the north and south poles.
They are restricted to lie in the hemispheres $V_{\pm \gamma}$ defined by
\be
V_{\gamma}= \{ \varpi : \, \Im(Z(\gamma)/\varpi) <0 \},
\label{lpmGMN2}
\ee
where $Z(\gamma)$ is the normalized central charge function on $H_3(X, \IZ)$,
\be
\label{defZ}
Z(\gamma) \equiv \frac{q_\Lambda z^\Lambda- p^\Lambda F_\Lambda(z)}{\sqrt{K(z,\bz)}}
\ee
with $K(z,\bz)=-2\Im (\bz^\Lambda F_\Lambda(z))$.
For convenience, we can fix the position of the rays at the middle of the hemispheres
\be
\ellg{\gamma}= \{ \varpi :\,  Z(\gamma)/\varpi \in \I\IR^{-} \} .
\label{rays}
\ee
Then we cover $\CP$ by four patches:
the first patch $\cU_+$ surrounds the north pole
and extends along the rays $\ellg{\pm \gamma}$ down to the equator. The second patch $\cU_-$ surrounds
the south pole and similarly  extends halfway along $\ellg{\pm \gamma}$,  with a non-vanishing
intersection with $\cU_+$. The rest of $\CP$ consists of two connected
parts covered by two patches $\cU_0$ and $\cU_{0'}$,
which overlap with $\cU_+$ and $\cU_-$ but stay away from the contours $\ellg{\pm\gamma}$.

Thus, one has to specify transition functions corresponding to the overlaps
of $\cU_0$ and $\cU_{0'}$ with $\cU_\pm$. The transition function between
$\cU_+$ and $\cU_-$ can then be determined by composing the previous ones.
We suggest that the twistor space incorporating the effect of one D-instanton
is produced by the following set
\be
\label{gensymp}
\begin{split}
\hHij{+0}&=\hHij{+0'}=  \frac{\I}{2}\[ F(\xii{+})
+\Gg -\frac{1}{2}\, q_\Lambda p^\Lambda\, (\Gg')^2\] ,
\\
\hHij{-0}& =\hHij{-0'}=   \frac{\I}{2}\[ \bF(\xii{-})
-\Gg -\frac{1}{2}\, q_\Lambda p^\Lambda\, (\Gg')^2\]
\end{split}
\ee
and the only non-vanishing anomalous dimension is $\ci{+}_\alpha= \chi_X/(96\pi)$.
Here we introduced the function
\be
\Gg(\Xikl)=\frac{\I\hnkl}{4\pi^3}
\int_0^{-\I\infty} \frac{\Xi\,\de\Xi}{\Xikl^2-\Xi^2}\, \Li_2\left(e^{-2\pi \I\, \Xi} \right),
\label{funGg}
\ee
$\Gg'$ denotes its derivative,
and the argument $\Xikl$ is related to the arguments of the transition functions
through the transcendental equation
\be
\Xiij{\pm 0} \equiv
q_\Lambda\xii{\pm}^\Lambda+2\I p^\Lambda \txii{0}_\Lambda
=\Xikl \mp q_\Lambda p^\Lambda \Gg'(\Xikl) .
\label{Xarg}
\ee
The coefficients $\hnkl$ are supposed to coincide with the generalized
Donaldson-Thomas invariants and for vanishing $p^\Lambda$ can be related to
the genus zero Gopakumar--Vafa invariants.
The holomorphic prepotential encodes the tree level part of the hypermultiplet metric,
the anomalous dimension gives rise to the one-loop correction, and the function $\Gg$
incorporates the effect of D2-instanton of charge $\gamma$.
In the linear approximation, where one neglects the quadratic term in $\Gg$,
the transition functions \eqref{gensymp} coincide with the ones
proposed in \cite{Alexandrov:2008gh}. Here the proposal is extended to all orders.
In sections \ref{subsec_transBPSone} and \ref{subsec_symplec}
we show that this extension is favored by symplectic invariance.

The transition functions \eqref{gensymp} are designed in such way in order to produce
discontinuities in $\xii{0}^\Lambda, \txii{0}_\Lambda$ along the BPS rays
given by simple $\log\(1-e^{-2\pi\I \Xikl}\)$. Indeed, when one evaluates derivatives
entering the gluing conditions \eqref{QKgluing}, one should take into account that
this amounts to differentiate the instanton contribution in \eqref{gensymp} with
respect to $\Xiij{\pm 0}$, whereas the function $\Gg$ depends on $\Xikl$.
Therefore, one must use the relation \eqref{Xarg}, which ensures the following important property
\be
\frac{d}{d\Xiij{\pm 0}}\(\Gg \mp \frac{1}{2}\, q_\Lambda p^\Lambda\, (\Gg')^2\)
=\Gg'(\Xikl).
\label{derG}
\ee
The function $\Gg'$ has the required discontinuity when $\Xikl$ crosses the integration contour
and one can adjust the contour in such way that in the $\varpi$-plane, at least near the poles,
it goes along $\ellg{\pm\gamma}$.

This consideration shows that the quadratic term is needed in order to take into account
the difference in the arguments of the transition functions and of the function $\Gg$.
It could be avoided if one takes $\Gg$ be dependent directly of $\Xiij{\pm 0}$.
However, as will become clear later, this would spoil the symplectic invariance.
Therefore, $\Gg$ must depend on $\Xikl$ defined through \eqref{Xarg},
which will turn out to be given by a symplectic invariant combination of Darboux coordinates
(see \eqref{Xisymple} below).
As a result, the transition functions generate a twistor space carrying
a representation of the symplectic group and
allow an {\it exact} solution for the twistor lines
(almost, {\it i.e.} up to quadratic terms appearing in the relation between the NS axion
and the parameters of the twistor lines)
coinciding with the linear approximation found in \cite{Alexandrov:2008gh}.

In order to present this solution, it will be convenient to work in the patch $\cU_0$,
where we expect the presence of symplectic invariance \cite{Alexandrov:2008nk}. Besides, it is convenient
to define the following combinations
\be
\label{defrho}
\rho_\Lambda\equiv  -2\I \txii{0}_\Lambda ,
\qquad
\tilde\alpha\equiv 4\I  \ai{0} + 2 \I \txii{0}_\Lambda \xii{0}^\Lambda.
\ee
Then we claim that the transition functions \eqref{gensymp} lead to
the following twistor lines
\bse
\label{xiqlineB}
\bea
\label{xiqlineB2}
\xii{0}^\Lambda &=& \zeta^\Lambda + \cR \left(
\varpi^{-1} z^\Lambda - \varpi \, \bz^\Lambda\right) +
\frac{\hnkl}{8\pi^2}\,p^\Lambda\, \Ikl^{(1)}(\varpi) ,
\\
\label{txiqlineB2}
\rho_\Lambda&=&
\tzeta_\Lambda
+\cR \left( \varpi^{-1} F_\Lambda - \varpi \, \bF_\Lambda \right)
+\frac{\hnkl}{8\pi^2}\, q_\Lambda\, \Ikl^{(1)}(\varpi) ,
\\
\label{txifqlineB2}
\tilde\alpha&=& \sigma
+\cR (\varpi^{-1} W-\varpi \,\bar W) +\frac{\I\chi_X}{24\pi} \log \varpi
+\frac{\I\hnkl}{\pi^2}
\(\varpi^{-1}\Wkl+\varpi\bWkl \)\Kkl
\nn\\
&& + \frac{\hnkl}{8\pi^2} \[\frac{1}{\pi \I}\,
\Ikl^{(2)}(\varpi)+\(\Thkl+\varpi^{-1}\Wkl-\varpi\bWkl\)\Ikl^{(1)}(\varpi) \] ,
\eea
\ese
where we defined the Type IIA fields
\bea
\label{xzeta}
&&\cR \,z^\Lambda =  Y^\Lambda,
\qquad
\zeta^\Lambda \equiv A^\Lambda ,
\qquad
\tzeta_\Lambda \equiv B_{\Lambda}+A^\Sigma \Re F_{\Lambda\Sigma}
+\frac{\hnkl}{2\pi^2}\,p^\Sigma \Im F_{\Lambda\Sigma}\,\Kkl ,
\\
\sigma \equiv & & \!\!\! - 2 B_\alpha  - \zeta^\Lambda \tzeta_\Lambda
+\(\zeta^\Lambda\zeta^\Sigma-\frac{\hnkl^2}{4\pi^4}\,p^\Lambda p^\Sigma\, \Kkl^2\)\Re F_{\Lambda\Sigma}
+\frac{\hnkl}{\pi^2}\, p^\Lambda \zeta^\Sigma  \Im F_{\Lambda\Sigma}\,\Kkl
+\frac{\hnkl^2}{4\pi^4}\,q_\Lambda p^\Lambda\, \Kkl^2,
\nonumber
\eea
and introduced
\be
\label{defWnon}
W(z) \equiv  F_\Lambda(z) \zeta^\Lambda - z^\Lambda \tzeta_\Lambda .
\ee
\be
\Wkl \equiv \cR \left( q_\Lambda z^\Lambda - p^\Lambda F_\Lambda(z) \right),
\qquad
\Thkl
\equiv q_\Lambda \zeta^\Lambda - p^\Lambda\tzeta_\Lambda,
\label{THg}
\ee
as well as
\be
\begin{split}
\Kkl\equiv &
\sum\limits_{m=1}^{\infty} \frac{1}{m} \sin\(2\pi m \Thkl\)\,
K_0\(4\pi m |\Wkl|\) ,
\\
\Ikl^{(\nu)}(\varpi)
\equiv &
\sum_{m=1}^{\infty} \sum_{s=\pm 1} \frac{s^\nu}{m^\nu}\,
e^{-2\pi \I s m\Thkl }
\int_{0}^{\infty}\frac{\d t}{t}\, \frac{t-\epskl s\I \varpi}{t+\epskl s\I\varpi}\,
e^{-2\pi m|\Wkl|( t^{-1}  +t )}  ,
\end{split}
\label{IKg}
\ee
with $\epskl =e^{-\I\arg \Wkl}$. Here $\cR$ is a real field which can be traded for the four-dimensional
dilaton $\phi$ coinciding with the contact potential.
Their relation is given below in \eqref{phiinstfull}.
In the patch $\cU_{0'}$ the twistor lines are given by the same formulae, although their
analytic continuation to $\cU_0$ does not coincide with \eqref{xiqlineB} due to the
discontinuities picked up by $\Ikl^{(\nu)}(\varpi)$.
In $\cU_\pm$ the twistor lines can be obtained by applying the gluing conditions \eqref{QKgluing}
with $\hHij{ij}$ from \eqref{gensymp}.

Since the twistor lines satisfy all gluing conditions by construction, to prove that
they follow from \eqref{QKgluing}, it is sufficient to show that they are regular everywhere
except for the singularities allowed in \eqref{txiqline}.
The regularity in the patches $\cU_0$ and $\cU_{0'}$ is evident because the only singularities of
$\Ikl^{(\nu)}(\varpi)$ are two cuts from $\varpi=0$ to $\varpi=\infty$, which belong to $\cU_+\cup \cU_-$.
Thus, it remains to check the analytic structure in the patches
$\cU_\pm$.

For this purpose, note that, due to \eqref{derG},
the analytic structure (besides simple poles at $\varpi=0,\infty$) of $\xii{\pm}^\Lambda$
and $\txii{\pm}_\Lambda$ is determined by the following combination\footnote{$\txii{+}_\Lambda$
contains additional non-trivial term $F_\Lambda(\xii{+})/(2\I)$. But since we prove that $\xii{+}^\Lambda$
is meromorphic in $\cU_+$, this term is also meromorphic.}
\be
\Ikl^{(1)}(\varpi)
\pm 4\[\int_0^{-\I\infty} \frac{\Xikl\,\de\Xi}{\Xikl^2-\Xi^2}\, \log\left(1-e^{-2\pi \I\, \Xi} \right)
+\frac{\pi}{12\I\Xikl}\].
\label{an_combin}
\ee
To analyze it, we need to know the dependence of $\Xikl$ on $\varpi$.
It can be found from \eqref{derG} and \eqref{Xarg}, which imply
\be
\Xikl=q_\Lambda\xii{0}^\Lambda-p^\Lambda\rho_\Lambda.
\label{Xisymple}
\ee
Using the explicit solution \eqref{xiqlineB}, one therefore finds
\be
\Xikl=\Thkl+\varpi^{-1}\Wkl-\varpi\bWkl
\label{resXig}
\ee
and thus $\Xikl$ does not contain instanton corrections.
Now it is easy to check that although both $\Ikl^{(1)}(\varpi)$ and the integral over $\Xi$
have two cuts starting from $\varpi=0$ ($\varpi=\infty$), their discontinuities cancel each other.
As a result, \eqref{an_combin} defines a meromorphic function in the neighborhood of the north (south) pole.

Finally, it remains to analyze $\alpha^{[\pm]}$.
Using \eqref{defrho} and the gluing conditions \eqref{QKgluing}, one obtains
\be
4\I \alpha^{[\pm]}=\tilde\alpha-2\I \xii{\pm}^\Lambda\txii{\pm}_\Lambda
\pm \(2\Gg-\Xikl \Gg' \).
\label{alphas}
\ee
Again, it is easy to check that the discontinuities of the last term cancel the discontinuities
in $\tilde\alpha$ generated by the terms in the last line of \eqref{txifqlineB2}.
Besides, the simple poles at $\varpi=0$ ($\varpi=\infty$), which are present in \eqref{txifqlineB2},
are all canceled by the second term in \eqref{alphas}.
Thus, the only singularity of $\alpha^{[\pm]}$ is given by the simple logarithmic term $\log\varpi$,
in accordance with \eqref{txiqline}.
This completes the proof of our solution \eqref{xiqlineB}. In appendix \ref{ap_twistline} we also
give a direct derivation of this solution from the integral representation \eqref{txiqline}.

The contact potential $\Phi_{\rm A/B}=\phi$ corresponding to the solution \eqref{xiqlineB} coincides
with the result found in the linear instanton approximation
and is given by
\be
e^{\phi} = \frac{\cR^2}{4}\, K(z,\bz)+\frac{\chi_X}{192\pi}
+\frac{ \hnkl}{4\pi^2}\,\sum\limits_{m> 0}
\frac{|\Wkl|}{m}\, \cos\(2\pi m \Thkl\)
K_1(4\pi m| \Wkl|) .
\label{phiinstfull}
\ee
Through \eqref{Knuflat} this result encodes the K\"ahler potential on $\cZ$.

As we claimed above, the contact twistor lines \eqref{xiqlineB} and the contact potential \eqref{phiinstfull}
are compatible with the action of the symplectic group. Indeed, if one simultaneously transforms
the type IIA fields and the charge vector $\gamma=(q_\Lambda,p^\Lambda)$, then
$\xii{0}^\Lambda$ and $\rho_\Lambda$ form a symplectic vector, whereas $\tilde\alpha$, $\phi$
and the variable $\Xikl$ are symplectic invariants.

\subsubsection{Transition functions through BPS rays}
\label{subsec_transBPSone}

It is instructive to find the transition function between the patches $\cU_0$
and $\cU_{0'}$. Although they do not intersect, we can define the corresponding transition function
through the composition law of elementary contact transformations. In the particular case,
where the transition functions $\hHij{ij}$ are independent of $\alpha^{[j]}$,
this law and the inverse transform are given by
\be
\begin{split}
\hHij{ij}&=\hHij{ik}+\hHij{kj}+\p_{\xii{k}^\Lambda}\hHij{kj}\p_{\txii{k}_\Lambda}\hHij{ik},
\\
\hHij{ji}& = - \hHij{ij}+\p_{\xii{i}^\Lambda}\hHij{ij}\p_{\txii{j}_\Lambda}\hHij{ij},
\end{split}
\label{compandinv}
\ee
where, if necessary, the domain of definition of transition functions is extended by means of
analytical continuation. Combining these equations, one finds
\be
\hHij{ij}=-\hHij{ki}+\hHij{kj}+\p_{\xii{k}^\Lambda}\(\hHij{ki}-\hHij{kj}\)\p_{\txii{k}_\Lambda}\hHij{ki}.
\label{comp}
\ee

The latter equation can already be applied to our twistor space. Let us specialize it for
$i=0,\ j=0',\ k=+$. In fact, one can introduce two transition functions relating $\cU_0$
and $\cU_{0'}$ because there are two inequivalent ways to perform the necessary
analytical continuation, either through $\ellg{\gamma}$ or through $\ellg{-\gamma}$. We attach the corresponding
index to $\hHij{00'}$ to distinguish these two cases.
Then it is easy to show that
\be
\hHijg{00'}{\pm}=\frac{\I}{2}\(\pm\Ggss{\pm\gamma}-\hf\,q_\Lambda p^\Lambda (\Ggss{\pm\gamma}')^2\) ,
\label{transell}
\ee
where
\be
\Ggs(\Xikl)=\frac{\hnkl}{(2\pi)^2}\,
\Li_2\left(e^{-2\pi \I \Xikl} \right)
\label{prepH}
\ee
and $\Xikl$ is expressed through the arguments of the transition functions by means of 
the following condition
\be
\Xiij{00'} \equiv
q_\Lambda\xii{0}^\Lambda+2\I p^\Lambda \txii{0'}_\Lambda
=\Xikl - q_\Lambda p^\Lambda \Ggs'(\Xikl) .
\label{Xargsim}
\ee

The transition functions \eqref{transell} can be considered as describing
a simplified version of our twistor space.
This simplified version appears if one
shrinks $\cU_\pm$ along the contours $\ellg{\pm\gamma}$ and reduce
them to small disks around the north and south poles of $\CP$, as shown on the right
of Fig. \ref{fig_sphere}. As a result, the two patches, $\cU_0$
and $\cU_{0'}$, have two disconnected common boundaries and the
transition functions through them are given by $\hHijg{00'}{\pm}$.
Such description is a very convenient setup to perform calculations
and it will be used in the next subsection to generalize the present construction
to the case of several D-instantons.

In \cite{Alexandrov:2008gh} a proposal for the exact contact transformations
through the BPS rays has been advocated.
In our notations it says that $\hHijg{00'}{\pm}=\pm\frac{\I}{2}\,\Ggss{\pm\gamma}(\Xiij{00'})$.
The result \eqref{transell}
coincides with that proposal only in the linear approximation.
At higher orders the difference is twofold.
First, the transition functions \eqref{transell} contain an additional quadratic term
proportional to the non-invariant combination of charges $q_\Lambda p^\Lambda$.
Second, the argument of $\Ggs$ does not coincide with the combination
of Darboux coordinates $\Xiij{00'}$. As we explained below \eqref{derG}, these two differences are
related and originate from the requirement that the discontinuities along the BPS rays
are simple logarithmic functions of the symplectic invariant $\Xikl$.
In turn, this is needed for symplectic invariance and to ensure that the linear
approximation is exact. It must be exact for any single charge $\gamma$ because,
as mentioned in the Introduction, an appropriate symplectic transformation maps it
to the form $(q'_\Lambda,0)$ corresponding to a D-brane wrapping an A-cycle,
which does possess this property.
On the other hand, for the linear approximation to be exact, it is necessary
that the argument of $\Ggs$ does not receive instanton corrections. This is true
for $\Xikl$ \eqref{resXig}, but it is not true for $\Xiij{00'}$.
Thus, we may claim that the proposal of \cite{Alexandrov:2008gh} is not consistent
with symplectic invariance, whereas our proposal is consistent, as follows from the explicit
results for the contact twistor lines. An additional argument supporting this statement
and formulated directly at the level
of transition functions will be given in section \ref{subsec_symplec}.

Notice that although $\Ggs$ coincides with the discontinuity of $\Gg$, this is not true for the whole
transition function: $\hHijg{00'}{\pm}$ cannot be obtained as a discontinuity of $\hHij{+0}$
due to the quadratic term. The correct way to obtain \eqref{transell} is by means of the composition rule
\eqref{comp} as described above.

\subsection{Inclusion of all instantons}
\label{subsec_all}

\subsubsection{Transition functions through BPS rays and twistor lines}
\label{subsec_alltr}

Let us include several instantons into consideration.
The starting point will be the transition function \eqref{transell}
through a BPS ray representing the contribution of one instanton.
Every instanton gives rise to two such rays on $\CP$, defined in \eqref{rays}
by the phase of its central charge.
Following \cite{Gaiotto:2008cd,Alexandrov:2008gh}, we assume that
across every BPS ray the Darboux coordinates experience contact transformations
generated by functions \eqref{transell}.

More precisely, let $\{ \gamma_a\}_{a=1}^N$ be a set of charges under consideration.
We assume that the phases of their central charges $Z(\gamma_a)$ are all different.
Then the BPS rays $\ellg{\pm \gamma_a}$ split $\CP$ into $2N$ sectors (see Fig. \ref{fig_many}).
For convenience we introduce additional $N$ charges, which are the opposite of the initial ones,
and we order all $2N$ charges, labeled by $i=1,\dots,2N$, in accordance
with decreasing of the phase of $Z(\gamma_i)$.
Furthermore, the sector bounded by $\ellg{\gamma_{i-1}}$ and $\ellg{\gamma_{i}}$
will be denoted by $\cU_i$ and we define
\be
\Xiijg{ij}{k} \equiv
q_{k,\Lambda}\xii{i}^\Lambda+2\I p_k^\Lambda \txii{j}_\Lambda.
\label{Xarggem}
\ee
With these definitions the transition functions through the BPS rays are
\be
\hHij{i\,i+1}=\frac{\I}{2}\(\Ggss{\gamma_i}-\hf\,q_{i,\Lambda} p_i^\Lambda (\Ggss{\gamma_i}')^2\) ,
\label{transellg}
\ee
where $\Ggss{\gamma_i}(\Xigi{i})$ with $\Xigi{i}\equiv\Xiijg{ii}{i}$ is defined in \eqref{prepH}.
They generate the contact transformations expressed by the following relations
between the contact Darboux coordinates
\be
\xii{j}^\Lambda=\xii{i}^\Lambda+\sum_{k=i}^{j-1} p_k^\Lambda\Ggss{\gamma_k}',
\qquad
-2\I\txii{j}_\Lambda=-2\I\txii{i}_\Lambda+\sum_{k=i}^{j-1} q_{k,\Lambda}\Ggss{\gamma_k}'.
\label{glumany}
\ee

\EPSFIGURE{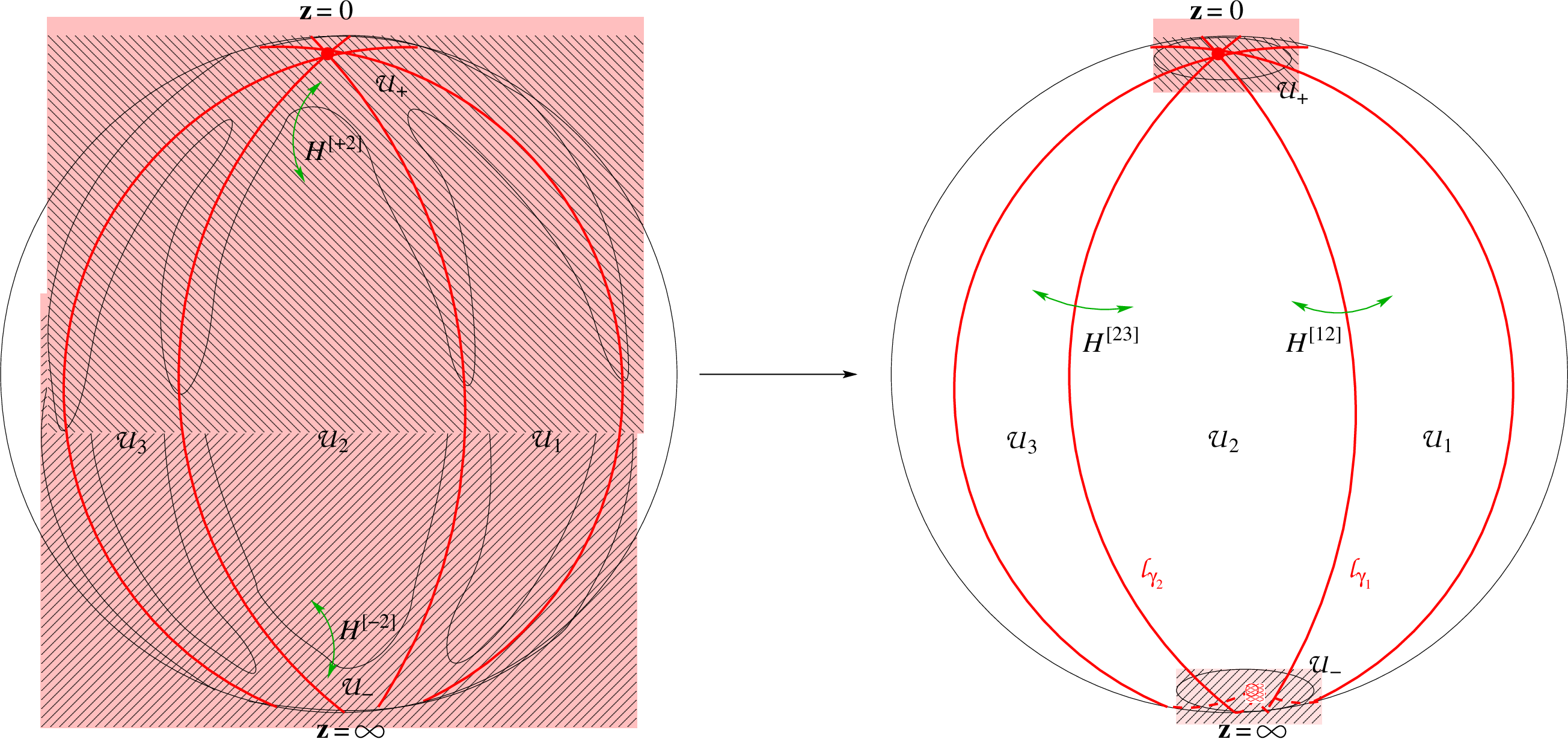,height=7.5cm}{Coverings of $\CP$ in the presence of several instantons. The
left picture defines a covering consisting of two patches around the poles and extending along
the BPS rays and $2N$ patches filling the remaining holes, where $N$ is the number of different charges.
The right picture is a limit of the left one and is more suitable for the analysis of the twistor
space. \label{fig_many}}

Although the gluing conditions \eqref{glumany} are linear in the functions $\Ggss{\gamma_i}$,
the resulting twistor space is much more complicated than the twistor space with a single instanton
presented in section \ref{subsec_single}. In particular, the linear instanton approximation
is not exact anymore. The easiest way to see this is to consider $\Xigi{i}$ in this approximation.
Substituting the results \eqref{xiqlineB}, where the instanton contributions should contain
the sum over all charges, into the definition \eqref{Xarggem}, one obtains \eqref{resXig}
plus the following contribution
\be
\frac{1}{8\pi^2}\sum_j \hng{j}\<\gamma_i,\gamma_j\> \Ig{j}^{(1)}(\varpi),
\label{resXmany}
\ee
where
\be
\<\gamma_1,\gamma_2\>=q_{1,\Lambda}p_2^\Lambda-q_{2,\Lambda}p_1^\Lambda
\label{scpr}
\ee
defines a symplectic invariant scalar product on the lattice of charges.
Thus, the instantons affect the variables, which the transition functions depend on.
As a result, their contributions propagate further and destroy the one-charge exact solution.

An exact solution can nevertheless be constructed by perturbative approach
described in appendix \ref{ap_pertsol}. For $\xi^\Lambda,\txi_\Lambda$,
it was analyzed in \cite{Gaiotto:2008cd}
where similar integral equations for the twistor lines of a HK space appeared.
It is easy to see that it can be written as
\be
\label{exline}
\begin{split}
\xii{i}^\Lambda &= \zeta^\Lambda + \cR \left(
\varpi^{-1} z^\Lambda - \varpi \, \bz^\Lambda\right) +
\frac{1}{8\pi^2}\sum_j \hng{j} p_j^\Lambda \Igg{j}(\varpi) ,
\\
-2\I\txii{i}_\Lambda &=
\tzeta_\Lambda
+\cR \left( \varpi^{-1} F_\Lambda - \varpi \, \bF_\Lambda \right)
+\frac{1}{8\pi^2}\sum_j \hng{j} q_{j,\Lambda}\Igg{j}(\varpi)  ,
\end{split}
\ee
where $\varpi\in \cU_i$,
\be
\Igg{i}(\varpi)=\int_{\ellg{\gamma_i}}\frac{\d \varpi'}{\varpi'}\,
\frac{\varpi+\varpi'}{\varpi-\varpi'}\,
\log\(1-e^{-2\pi \I \Xigi{i}(\varpi')}\),
\label{newfun}
\ee
and $\Xigi{i}(\varpi)$ should be found as a solution of the following system of equations
\be
\label{eqXigi}
\Xigi{i}(\varpi)=\Thkli{i}+\varpi^{-1}\Wkli{i}-\varpi\bWkli{i}+
\frac{1}{8\pi^2}\sum_{j\ne i} \hng{j}\<\gamma_i,\gamma_j\> \int_{\ellg{\gamma_j}}\frac{\d \varpi'}{\varpi'}\,
\frac{\varpi+\varpi'}{\varpi-\varpi'}\,
\log\(1-e^{-2\pi \I \Xigi{j}(\varpi')}\).
\ee
The latter equations encode all non-trivialities of the problem. They can be analyzed
perturbatively and their solution, represented by a set of variables $\Xigi{i}(\varpi)$,
contains all orders of the instanton expansion.

One can also obtain a similar representation for the Darboux coordinate $\alpha$
which is a bit more complicated.
By appropriately adjusting the considerations in appendix \ref{ap_twistline}, one finds
the following result, written again for the combination $\tilde\alpha$ defined
in \eqref{defrho},
\bea
\tilde\alpha^{[i]}&=& \sigma
+\cR (\varpi^{-1} W-\varpi \,\bar W) +\frac{\I\chi_X}{24\pi} \log \varpi
-\frac{1}{4\pi^2}\sum_j \hng{j}\(\varpi^{-1}\Wg{j}+\varpi\bWg{j} \)\Igg{j}(0)
\nn\\
&& + \frac{1}{8\pi^2}\sum_j \hng{j} \[\frac{\I}{\pi }
\int_{\ellg{\gamma_j}}\frac{\d \varpi'}{\varpi'}\,
\frac{\varpi+\varpi'}{\varpi-\varpi'}\,
\Li_2\(e^{-2\pi \I \Xigi{j}(\varpi')}\)
+\(\Thg{j}+\varpi^{-1}\Wg{j}-\varpi\bWg{j}\)\Igg{j}(\varpi) \]
\nn \\
&& +\frac{1}{64\pi^2}\sum_{j\ne k} \hng{j}\hng{k}\<\gamma_j,\gamma_k\>
\int_{\ellg{\gamma_j}}\frac{\d \varpi'}{\varpi'}\,
\frac{\varpi+\varpi'}{\varpi-\varpi'}\,\log\(1-e^{-2\pi \I \Xigi{j}(\varpi')}\)
\Igg{k}(\varpi').
\label{manyalpha}
\eea

\subsubsection{Transition functions to the poles}
\label{subsec_allrev}

To complete the definition of the twistor space,
we need to provide two additional transition functions to the patches $\cU_\pm$
around the poles of $\CP$. We can find them requiring that they lead to regular contact
Darboux coordinates in these patches, {\it i.e.} that they cancel singularities of
\eqref{exline} and \eqref{manyalpha}.
The main complication comes from the fact that the variables $\Xigi{i}(\varpi)$ defined by \eqref{eqXigi}
acquire branch cuts along the BPS rays $\ellg{\gamma_j},\ j\ne i$.\footnote{These variables
possess interesting monodromies around the north and south poles. Since every charge $\gamma_a$
appears in the sum together with its opposite $-\gamma_a$, the total contribution of this pair of charges
to the monodromy is linear in $\Xigi{a}$. As a result, going around the poles in the clockwise
direction, one obtains the following transformation
$$
\Xigi{i}\longrightarrow \Xigi{i}-\sum_a \hng{a}\<\gamma_i,\gamma_a\>\Xigi{a} \quad {\rm mod}\ n\in \IZ.
$$
Taking into account that $\hng{a}$ are expected to be integers, we see that the monodromy gives
rise
to a linear combination of $\Xigi{a}$ with integer coefficients and can be represented as
a linear map on the charge lattice:
$$
\( q_\Lambda \atop p^\Lambda\) \longrightarrow \( q_\Lambda \atop p^\Lambda\)
-\sum_{\gamma'\in \Gamma/\IZ_2} n_{\gamma'}\<\gamma,\gamma'\> \( q'_\Lambda \atop p^{'\Lambda}\),
$$
where $\Gamma/\IZ_2$ is the charge lattice modulo identification $\gamma\leftrightarrow -\gamma$.
}
As a result, the simple sum of one-charge contributions \eqref{gensymp} does not work anymore because,
canceling the singularities of the initial Darboux coordinates
by means of integrals, it introduces other singularities through
the dependence of $\Xigi{i}(\varpi)$. Therefore, a more elaborated construction is required.

The correct transition functions to the patch $\cU_+$ may be given in terms of $\Xipgi{i}\equiv \Xiijg{++}{i}$
and the functions $\Sgi{i}(\Xi)$ defined by the following system of integral equations
\be
\Sgi{i}(\Xi)=\frac{1}{8\pi^2}\sum_j \hng{j}\<\gamma_i,\gamma_j\>\[
\int_0^{-\I\infty}\!\!\!\! \frac{2\,\Xi\,\de\Xi'}{\Xipgi{j}\Xi-\Xipgi{i}\Xi'}\,
\log\left(1-e^{-2\pi \I\( \Xi'-\Sgi{j}(\Xi')\)} \right)+\cF_j\],
\label{defSF}
\ee
where the functions $\cF_j$ satisfy
\be
\begin{split}
\cF_j=& \frac{\pi \I}{6\Xipgi{j}}+\frac{1}{2\pi^2}\sum_k \hng{k}\<\gamma_j,\gamma_k\>
\int_0^{-\I\infty}\!\!\Dlog{j}
\int_0^{-\I\infty}\!\! \frac{\Xi'\Dlog{k}'}{(\Xipgi{k}\Xi-\Xipgi{j}\Xi')^2}
\\
& +\frac{1}{16\pi^2}\sum_{k\ne l} \frac{\hng{k}\hng{l}}{\hng{j}}\,\<\gamma_k,\gamma_l\>
\cF_k\,\frac{\p\cF_l}{\p\Xipgi{j}},
\end{split}
\label{regfun}
\ee
and we introduced for convenience the measure
\be
\Dlog{}= \log\left(1-e^{-2\pi \I \( \Xi-\Sgi{}(\Xi) \)} \right)\de\Xi.
\label{measure}
\ee
Similarly to the equations for $\Xigi{i}$ \eqref{eqXigi}, they can be solved perturbatively
by expanding in powers of instantons, which is also equivalent to the expansion in
invariants $\hng{i}$.\footnote{Note that $\hng{j}$ in the denominator of the last term in \eqref{regfun}
is canceled after substituting ${\p_{\Xipgi{j}}\cF_l}$.}

Since $\Xipgi{i}$ is by definition meromorphic around the north pole,
the singularities in the above expressions may appear only from the integrals.
However, the denominator in \eqref{defSF} and \eqref{regfun} can be approximated by
$\varpi^{-1}\(\Wg{j}\Xi-\Wg{i}\Xi'\)$ which does not vanish if the phases of $\Wg{i}$
and $\Wg{j}$ are different. Thus, both $\cF_j$ and $\Sgi{i}(\Xi)$ are regular functions
in the patch $\cU_+$. However, the latter function possesses an important property that
for $\Xi=\Xipgi{i}$ it gives rise to
\be
\Sgi{i}(\Xipgi{i})=
\frac{1}{8\pi^2}\sum_j \hng{j}\<\gamma_i,\gamma_j\>
\derG{j},
\label{derGx}
\ee
where
\be
\derG{j}=2\int_0^{-\I\infty}\!\! \frac{\de\Xi}{\Xipgi{j}-\Xi}\,
\log\left(1-e^{-2\pi \I\( \Xi-\Sgi{j}(\Xi)\)} \right)+\cF_j
\label{instXi}
\ee
has a branch cut starting from $\varpi=0$. If we impose the relation
\be
\Xigi{i}=\Xipgi{i}-\Sgi{i}(\Xipgi{i}),
\label{relXXp}
\ee
then the discontinuity along this cut is the same as the discontinuity of the function
$\Igg{i}$ \eqref{newfun}. This is consistent with the regularity of $\Xipgi{i}$ in $\cU_+$
because the relation \eqref{relXXp} ensures that the corresponding singularities of
$\Sgi{i}(\Xipgi{i})$ cancel those of $\Xigi{i}$.
The functions $\cF_j$ are not really important since they are regular and can be canceled
by appropriate (although complicated) gauge transformation.

Now we are ready to define the transition functions. They are given by
\be
\begin{split}
\hHij{+i}=&
\frac{\I}{2}\, F(\xii{+})
-\frac{1}{16\pi^3}\sum_j\hng{j}
\int_0^{-\I\infty}\!\! \frac{\de\Xi}{\Xipgi{j}-\Xi}\,
\Li_2\left(e^{-2\pi \I\(\Xi-\Sgi{j}(\Xi)\)} \right)
\\
&
-\frac{\I}{(16\pi^2)^2}\sum_{j\ne k} \hng{j}\hng{k} q_{j,\Lambda}p_k^\Lambda \derG{j}\derG{k}
-\frac{\I}{16\pi^2}\sum_{j} \hng{j}
\int_0^{-\I\infty}\!\! \frac{\Dlog{j}}{\Xipgi{j}-\Xi}\, \Sgi{j}(\Xi) .
\end{split}
\label{transpole}
\ee
In appendix \ref{ap_transpole} we verify that they indeed define regular Darboux coordinates in the
patch $\cU_+$ and they are consistent with the transition functions through the BPS rays \eqref{transellg}.
In a similar way one can construct the transition functions to the patch $\cU_-$.
For this it is sufficient to replace $F(\xii{+})$ by $\bF(\xii{-})$, $\Xipgi{i}$ by $\Ximgi{i}$,
and to flip the sign of $\hng{j}$ in all above equations.

Once we know the full set of transition functions, it is possible to compute the contact
potential coinciding with the four-dimensional dilaton.
This calculation uses the property
\be
\xii{\pm}(\varpi)=\xii{i}(\varpi)+O(\varpi^{\pm 2}),
\label{relXXexp}
\ee
which can be established from \eqref{derivH}, \eqref{instXi} and \eqref{regfun}.
Then the general formula for (the constant part of) the contact potential \eqref{contpotconst}
gives the following result
\be
e^{\phi} = \frac{\cR^2}{4}\, K(z,\bz)+\frac{\chi_X}{192\pi}
-\frac{ \I}{32\pi^2}\sum\limits_{j} \hng{j}
\int_{\ellg{\gamma_j}}\frac{\d \varpi}{\varpi}\,
\( \varpi^{-1}\Wg{j} -\varpi\bWg{j}\)
\log\(1-e^{-2\pi \I \Xigi{j}(\varpi)}\).
\label{phiinstmany}
\ee

\subsubsection{Restriction to mutually local states}
\label{subsec_local}

Let us note that if we restrict our attention to a sector in the charge lattice
where all charges satisfy the following condition
\be
\label{restr}
\<\gamma_i,\gamma_j\>=0,
\ee
all above mentioned complications, arising in the presence of several charges,
disappear. Indeed, substituting it into \eqref{eqXigi}, one finds that $\Xigi{i}(\varpi)$ do not
contain instanton corrections as in the single-charge case.
Moreover, this is true for all $\Xiijg{jj}{i}$ which turn out to be independent of the label $j$,
\be
\Xiijg{jj}{i}= \Thg{i}+\varpi^{-1}\Wg{i}-\varpi\bWg{i}.
\label{resX}
\ee
As a result, the linear instanton approximation becomes again exact.
The contact twistor lines are given by \eqref{exline}, where the function $\Igg{}(\varpi)$
may be replaced by the function $\textstyle \hf\, \Ikl^{(1)}(\varpi)$ \eqref{IKg}\footnote{More precisely,
\eqref{resX} ensures that $\Igam{\gamma}(\varpi)-\Igam{-\gamma}(\varpi)=\Ikl^{(1)}(\varpi)$.
Combining the contributions of charges $\gamma$ and $-\gamma$, one obtains the simple factor
$\textstyle \hf$.}, and by \eqref{manyalpha} where the last term vanishes and the same replacement
may be done.
Similarly, the transition functions \eqref{transpole} considerably simplify
and have the same structure \eqref{gensymp} as for single charge: the last term is absent
and the third term coincides with the product of derivatives of the second term.

Thus, the condition \eqref{restr} gives rise to an exactly solvable sector.
The exact twistor lines can be used to extract the metric on the HM moduli space
affected by D2-instantons with charges satisfying this restriction,
which we call by ``mutually local states".
Although this procedure is straightforward,
we do not present here the final result since it has a quite involved form.

It is clear that a particular set of mutually local states is given
by D2 branes wrapping only A-cycles which all have $p^\Lambda=0$.
This is a maximal possible set of such states and any other maximal set can be obtained
by a symplectic transformation.\footnote{For the universal hypermultiplet
\cite{Strominger:1997eb,Ketov:2001ky,Antoniadis:2003sw,Davidse:2005ef,Alexandrov:2006hx,Chiodaroli:2008rj}
this means that mutually local states can include only D-branes wrapping either A or B cycle.}
Due to this, it is not surprising that any such sector is exactly solvable.
This is just a consequence of such solvability for A-type D2-instantons.
Nevertheless, this still provides a non-trivial cross-check on our results.

\subsection{Discussion}

\subsubsection{Symplectic invariance}
\label{subsec_symplec}

Although the found twistor lines form a nice representation of the symplectic group, one can
ask whether the conditions of symplectic invariance can be imposed directly on
transition functions. Naively, one could expect that they should be invariant
under symplectic transformations. But for the functions $\hHij{\pm i}$ this is not true
even at the tree level due to the presence of the holomorphic prepotential.
As for the transition functions through the BPS rays \eqref{transellg},
they are not invariant due to the factor $q_{\Lambda} p^\Lambda$
in front of the additional quadratic term.

To understand what is going on, let us consider the gluing conditions \eqref{QKgluing}.
If in both patches $\cU_i$, $\cU_j$, the Darboux coordinates $\xi^\Lambda$, $-2\I\txi_\Lambda$
form a symplectic vector and $\tilde\alpha$ is invariant, the same property must hold
for the appropriate combinations of derivatives of the transition function $\hHij{ij}$.
However, it is difficult to convert this condition into a restriction on $\hHij{ij}$
itself. The reason is that the transition functions used so far relate
coordinate systems in two different patches and therefore depend also on coordinates
in these different patches. On the other hand, symplectic invariance holds only for quantities
defined in a single patch.

It is well known that generically one cannot write a symplectomorphism or a contact transformation
in terms of a function dependent only on initial coordinates. Let us nevertheless find the conditions
under which this might be possible.
If one writes the contact transformation (which induces a symplectomorphism in the
$(\xi,\txi)$-subspace) as
\be
\begin{split}
& \xii{j}^\Lambda=  \xii{i}^\Lambda-\p_{\txii{i}_\Lambda}\hhHij{ij},
\qquad
\txii{j}_\Lambda=\txii{i}_\Lambda+\p_{\xii{i}^\Lambda}\hhHij{ij},
\\
\alpi{j}& =\alpi{i}+\hhHij{ij}-\xii{i}^\Lambda\p_{\xii{i}^\Lambda}\hhHij{ij}
+\hf\,\p_{\txii{i}_\Lambda}\hhHij{ij}\p_{\xii{i}^\Lambda}\hhHij{ij},
\end{split}
\label{onepatchfun}
\ee
where $\hhHij{ij}$ is a function of $\xii{i}^\Lambda$ and $\txii{i}_\Lambda$,
and requires that the contact one-form $\cX$ \eqref{con1fo} is preserved,
one finds the following condition on $\hhHij{ij}$
\be
\p_{\xii{i}^\Lambda}\hhHij{ij}\, d\(\p_{\txii{i}_\Lambda}\hhHij{ij}\) =
\p_{\txii{i}_\Lambda}\hhHij{ij}\, d\( \p_{\xii{i}^\Lambda}\hhHij{ij}\).
\label{condsymple}
\ee

Assume for a moment that this condition is satisfied. Then the symplectic properties
of the contact Darboux coordinates $\xi^\Lambda$, $\txi_\Lambda$ require that
$(\p_{\txii{i}_\Lambda}\hhHij{ij},2\I\p_{\xii{i}^\Lambda}\hhHij{ij})$ is a symplectic vector,
whereas from the gluing condition on $\tilde\alpha$
\be
\tilde\alpha^{[j]} =\tilde\alpha^{[i]}+2\I\(2\hhHij{ij}-\xii{i}^\Lambda\p_{\xii{i}^\Lambda}\hhHij{ij}
-\txii{i}_\Lambda\p_{\txii{i}_\Lambda}\hhHij{ij}\)
\label{glutildealp}
\ee
one concludes that $\hhHij{ij}$ must be invariant under symplectic transformations.
The latter condition automatically solves also the former.
Thus, if it is possible to write the contact transformations between different patches
in terms of transition functions dependent on coordinates in one patch, symplectic
invariance simply requires the invariance of these functions.

Remarkably, the condition \eqref{condsymple} ensuring this possibility
is satisfied by functions for which all dependence on the contact Darboux coordinates
is through the symplectic invariant combination $\Xigi{i}$.
Thus, there is a class of contact transformations for which the generating functions can
be taken as in \eqref{onepatchfun}.
In particular, we can define the following generating functions
\be
\hhHij{i\,i+1}(\xii{i},\txii{i})=\frac{\I}{2}\,\Ggss{\gamma_i}(\Xigi{i}),
\label{transsimple}
\ee
which are explicitly symplectic invariant.
It is easy to see that they generate the same contact transformations
through the BPS rays as the ones generated by \eqref{transellg}.
This proves in another way that our construction respects symplectic invariance.

If one allows the transition functions to depend on several combinations $\Xiijg{ii}{j}$
associated with different charges $\gamma_j$, the condition \eqref{condsymple}
is not satisfied unless the charges are mutually local \eqref{restr}.
This is related to the phenomenon of wall crossing considered in the next subsection.
Besides, the non-invariance of the tree level part expressed through the holomorphic
prepotential is not in contradiction with symplectic invariance
because it appears in the transition functions to the patches $\cU_\pm$ only.
But the Darboux coordinates in these patches do not
form a representation of the symplectic group. Therefore, the above arguments
cannot be applied to $\hhHij{\pm i}$.

The representation \eqref{transsimple} is very nice since the non-invariant quadratic terms disappear and
all considerations become particularly simple. Note however the presence of a
quadratic term in the gluing condition for $\alpha$ \eqref{onepatchfun}, which
was absent in \eqref{Tfct}. Thus, we traded quadratic terms in the transition functions
for similar terms in the gluing conditions.
It is possible to get rid of them everywhere if one simultaneously trades $\alpha$
for $\tilde\alpha$ since the gluing condition \eqref{glutildealp} for the latter is linear.

\subsubsection{Wall crossing}
\label{subsec_wall}

In \cite{Gaiotto:2008cd} a physical explanation for the so called
wall-crossing formula \cite{ks} has been suggested in the context of $\cN=2$
supersymmetric gauge theories. It has been interpreted as a condition on the moduli
space metric to be continuous across the lines of marginal stability (LMS), where the BPS
spectrum of single-particle states is known to jump
(see e.g. \cite{Argyres:1995gd,Ferrari:1996sv,Bilal:1996sk}). The wall-crossing formula
relates the single instanton contribution on one side of the LMS
to the multi-instanton contribution on the other side.
This provides strong constraints on these contributions.

A similar phenomenon as LMS is known to take place in $\cN=2$ supergravity theories
and in \cite{Alexandrov:2008gh} it was suggested that the wall-crossing formula is
relevant also in this case expressing the condition of regularity of the metric on
the hypermultiplet moduli space. Here we would like to show that our construction
is consistent with the wall crossing.

The wall crossing condition requires that the  ``generalized Donaldson-Thomas invariants''
$\Omega(\gamma)$ defined in \cite{ks}
must satisfy
\be
\label{wc}
\prod^\ccwarrow_{\substack{\gamma=n \gamma_1+m \gamma_2\\m>0, n>0} }
U_{\gamma}^{\Omega^-(\gamma)} =
\prod^\cwarrow_{\substack{\gamma=n \gamma_1+m \gamma_2\\m>0, n>0} }
U_{\gamma}^{\Omega^+(\gamma)} ,
\ee
where $\Omega^{-}(\gamma)$ and $\Omega^{+}(\gamma)$ denote the value of $\Omega(\gamma)$ on
either side of the LMS where the phases of the central charges \eqref{defZ},
$Z(\gamma_1)$ and $Z(\gamma_2)$, align. Here
\be
\label{wcu}
U_{\gamma} \equiv \exp\left( \sum_{n=1}^{\infty} \frac{1}{n^2}\, e_{n\gamma} \right)
\ee
is a group element constructed from the generators $e_\gamma$ of the
following Lie algebra
\be
\label{algks}
\left[ e_{\gamma}, e_{\gamma'} \right] = (-1)^{ \<\gamma,\gamma'\>}
\<\gamma,\gamma'\> e_{\gamma+\gamma'} .
\ee
Except for the sign $(-1)^{ \<\gamma,\gamma'\>}$,
which can be absorbed into a redefinition of  $e_{\gamma}$ by a choice of
``quadratic refinement" \cite{Gaiotto:2008cd}, this is the algebra of infinitesimal
symplectomorphisms on the complex torus $ (\IC^\times)^{2N}$.

In our case the LMS appear because the order of the BPS rays, determined by the phases of
the central charges $Z(\gamma_i)$, is important.
Despite apparent commutativity of the contact transformations \eqref{glumany},
it is illusive because the exchange of two charges changes relations between
the arguments $\Xigi{i}$ of the functions entering the transformations.
To see this explicitly and to compare with the wall-crossing,
it is instructive to compute the composition of two transition functions
\eqref{transellg} associated with charges $\gamma_1$ and $\gamma_2$.
From the composition law \eqref{compandinv}, one finds
\be
\hHij{13}=\frac{\I}{2}\[\Ggss{\gamma_1}+\Ggss{\gamma_2}
-\hf\(q_{1,\Lambda}\Ggss{\gamma_1}'+q_{2,\Lambda}\Ggss{\gamma_2}'\)
\(p_1^\Lambda \Ggss{\gamma_1}'+p_2^\Lambda \Ggss{\gamma_2}'\)
+\hf\<\gamma_1,\gamma_2\>\Ggss{\gamma_1}'\Ggss{\gamma_2}'\].
\label{twoG}
\ee
From this result and using the relations \eqref{glumany} between the coordinates in different patches,
one can find the commutator of two transformations $U_{\gamma_i}$ generated by
\eqref{transellg} with charges $\gamma_1$ and $\gamma_2$. Expanding in powers of $\Ggss{\gamma_i}$,
at quadratic order one obtains
\be
U_{\gamma_2}^{-1}U_{\gamma_1}^{-1}U_{\gamma_2}^{\hphantom{1}} U_{\gamma_1}^{\hphantom{1}}
\approx -\<\gamma_1,\gamma_2\>\Ggss{\gamma_1}'\Ggss{\gamma_2}'+O(\Ggs^3).
\ee
It is clear that if we identify $e^{-2\pi\I\Xikl}$ with the elements $e_\gamma$, this commutation relation
coincides with the commutator \eqref{algks}, which is at the basis of the wall-crossing formula.
The construction from the previous subsection makes the comparison even more direct since
then the transition functions \eqref{transsimple} are identical to (the logarithm of)
the group elements \eqref{wcu}.

Finally, we note that the composition law \eqref{twoG} for two contact transformations through the BPS rays
can be easily generalized to the case of $n$ charges:
\be
\hHij{1\, n+1}=\frac{\I}{2}\[\sum_i \Ggss{\gamma_i}
-\hf\(\sum_i q_{i,\Lambda}\Ggss{\gamma_i}'\)
\(\sum_i p_i^\Lambda \Ggss{\gamma_i}'\)
+\hf\sum_{i<j} \<\gamma_i,\gamma_j\>\Ggss{\gamma_i}'\Ggss{\gamma_j}'\].
\label{manyG}
\ee
On the other hand, the above results imply that the contact transformations generated by the mutually local
states of section \ref{subsec_local} are commutative, which means that the LMS phenomenon does not occur
in such sector.

\acknowledgments
The author is grateful to Sylvain Ribault for valuable discussions
and especially to Boris Pioline and Stefan Vandoren for careful reading of the manuscript
and important comments.
This research is supported by CNRS.

\appendix

\section{Details on the twistor description of HK spaces}

\subsection{Solution for the Lagrangian}
\label{ap_Ksol}

Here we want to prove the formula \eqref{defL} for the Lagrangian, {\it i.e.}
that it ensures the conditions \eqref{condLang}.
For $\cL$ given by this formula, by simple manipulations, one can establish the following relations
\be
\begin{split}
\cL_{v^I}&=\p_{v^I} u^J\oint_C \frac{\de\zeta}{2\pi \I\, \zeta^2}\, H_J
-\p_{v^I} \bu^J\oint_C \frac{\de\zeta}{2\pi \I}\, H_J ,
\\
\cL_{\bv^I}&=\p_{\bv^I} u^J\oint_C \frac{\de\zeta}{2\pi \I\, \zeta^2}\, H_J
-\p_{\bv^I} \bu^J\oint_C \frac{\de\zeta}{2\pi \I}\, H_J ,
\\
\cL_{x^I}&=\oint_C \frac{\de\zeta}{2\pi \I\, \zeta}\, H_J
+\p_{x^I} u^J\oint_C \frac{\de\zeta}{2\pi \I\, \zeta^2}\, H_J
-\p_{x^I} \bu^J\oint_C \frac{\de\zeta}{2\pi \I}\, H_J ,
\\
\cL_{\vrh_I}&=\frac{\I}{2}\oint_C \frac{\de\zeta}{2\pi \I\, \zeta}\, H^J
+\p_{\vrh_I} u^J\oint_C \frac{\de\zeta}{2\pi \I\, \zeta^2}\, H^J
-\p_{\vrh_I} \bu^J\oint_C \frac{\de\zeta}{2\pi \I}\, H^J,
\end{split}
\ee
where $\cL$ on the l.h.s. is considered as a function of
$v^I,\bv^I,x^I,\vrh_I$. Comparing this with the relations between derivatives
\be
\begin{split}
\left.\p_{v^I}\right|_{\bv,x,\vrh}&=\p_{v^I} u^J\p_{u^J}+\p_{v^I} \bu^J\p_{\bu^J} ,
\\
\left.\p_{\bv^I}\right|_{v,x,\vrh}&=\p_{\bv^I} u^J\p_{u^J}+\p_{\bv^I} \bu^J\p_{\bu^J},
\\
\left.\p_{x^I}\right|_{v,\bv,\vrh}&=\p_{x^I}+\p_{x^I} u^J\p_{u^J}+\p_{x^I} \bu^J\p_{\bu^J} ,
\\
\left.\p_{\vrh_I}\right|_{v,\bv,x}&=\p_{\vrh_I}+\p_{\vrh_I} u^J\p_{u^J}+\p_{\vrh_I} \bu^J\p_{\bu^J} ,
\end{split}
\ee
one immediately concludes that the Lagrangian $\cL$ satisfies the conditions
\eqref{condLang}.
It is equivalent to the previous perturbative results \cite{Alexandrov:2008ds}
due to the relation
\be
\oint_C \frac{\de\zeta}{2\pi \I\, \zeta}\(\etap^I H_I-\mup_I H^I\)
=(u^I-v^I)\p_{u^I}\cL+(\bu^I-\bv^I)\p_{\bu^I}\cL.
\ee

The fact that the complex structures must satisfy the algebra of the quaternions, e.g.
\be
J^+\, J^-=-\hf\( {\bf 1}+\I J^3 \)
 \quad \leftrightarrow \quad
(\omega^+)_{\alpha\gamma}\,K^{\gamma\bar \delta}\,(\omega^-)_{\bar \delta\bar \beta}
=K_{\alpha\bar \beta} ,
\label{relmetrics}
\ee
imposes restrictions on the inverse metric.
It is easy to find that it must be given by
\be
\begin{split}
& K^{\bu^I u^J} =
\CL_{\vrh_I\vrh_J}-\cL_{\vrh_I x^K}\cL^{x^K x^L} \cL_{x^L\vrh_J}-\CL^{x^I x^J}
-\I\[\CL^{x^I x^K}\CL_{x^K\vrh_J}-\CL_{\vrh_I x^K}\CL^{x^K x^J} \] ,
\\
& K^{\bu^I w_J} = -\CL^{x^I x^K}\CL_{x^K u^J}
-\I\[\CL_{\vrh_I u^J}-\CL_{\vrh_I x^K}\CL^{x^K x^L}\CL_{x^L u^J}\] ,
\\
& K^{\bw_I u^J} = -\CL_{\bu^I x^K}\CL^{x^K x^J}
+\I\[\CL_{\bu^I \vrh_J}-\CL_{\bu^I x^M}\CL^{x^M x^N}\CL_{x^N\vrh_J}\]  ,
\\
& K^{\bw_I w_J} = \CL_{\bu^I u^J}-\CL_{\bu^I x^K}\CL^{x^K x^L}\CL_{x^L u^J} .
\end{split}
\label{HKmetinv}
\ee
This result requires that the Lagrangian satisfies some additional constraints
in order to ensure that \eqref{HKmetinv} is indeed the inverse of \eqref{HKmet}.
These constraints are a consequence of the fact that $\cL$ is given by
a holomorphic function of only $2d+1$ variables and generalize the well known
constraints in the $\cO(2)$ case \cite{deWit:2001dj}, which were written recently
also in the presence of linear perturbations \cite{Alexandrov:2008ds}.

\subsection{Perturbative solution}
\label{ap_pertsol}

The integral equations \eqref{soltwist} and \eqref{txiqline} are a very
convenient starting point to find twistor lines by the perturbative method.
This approach can be applied if the transition functions can be
represented as
\be
\Hij{ij}=\sum_{n=0}^{\infty}\lambda^n\Hnij{n}{ij},
\ee
where $\Hnij{0}{ij}$ are all independent of $\mui{j}_I$ and $\lambda$ is an infinitesimal parameter.
Then at the zeroth approximation, $\eta^I$ are global $\cO(2)$ multiplets which can be viewed
as moment maps for the isometries along $\vrh_I$. At the next orders in the parameter $\lambda$ however
this structure is destroyed.
Expanding in $\lambda$, the integral equations \eqref{soltwist} lead to the following solution
\be
\etai{i}^I=\etaz^I+\sum_{n=1}^{\infty}\lambda^n \etaniI{n}{i}{I},
\qquad
\mui{i}_I=\frac{\I}{2}\,\vrh_I+\sum_{n=0}^{\infty}\lambda^n \muniI{n}{i}{I},
\ee
where $\etaz^I$ was defined in \eqref{omult} and
\be
\begin{split}
\etaniI{n}{i}{I} &=-\hf \sum_j\oint_{C_j} \frac{\de\zeta'}{2\pi \I\,\zeta'}\,
\frac{\zeta'^3+\zeta^3}{\zeta\zeta' (\zeta'-\zeta)}
\[ \Hn{n}^{[ij]I}+\cdots\],
\\
\muniI{n}{i}{I} &=\hf\sum_j\oint_{C_j} \frac{\de\zeta'}{2\pi \I\,\zeta'}\,
\frac{\zeta'+\zeta}{\zeta'-\zeta}\[ \Hij{ij}_{(n)I}+\cdots+
\etaniI{n}{i}{J}\Hij{ij}_{(0)IJ}\] ,
\end{split}
\ee
where the transition functions are considered as functions of
$\etaz,{\textstyle\frac{\I}{2}}\vrh+\munI{0}{[j]}$ and $\zeta'$,
and the dots correspond to other terms of $n$th order appearing in the expansion of $\Hij{ij}$,
which all depend only on $\etanI{k}{[i]}$, $\munI{k}{[j]}$, $k<n$.
Thus, this leads to a well defined iteration procedure.

\section{Derivation of contact twistor lines for single charge}
\label{ap_twistline}

In this appendix we demonstrate that the twistor lines \eqref{xiqlineB} satisfy the
integral equations \eqref{txiqline}.

First, we compute the functions entering the gluing conditions \eqref{QKgluing} and defined
in \eqref{Tfct}. Due to \eqref{derG}, one has
\be
\begin{split}
& T_{[+ 0]}^\Lambda=p^\Lambda \Gg'(\Xikl),
\qquad
\tilde{T}_\Lambda^{[+0]}=\frac{\I}{2}\(F_\Lambda(\xii{+})+q_\Lambda\Gg'(\Xikl)\),
\\
\tilde{T}^{[+0]}_\alpha & =\frac{\I}{2}\(-F(\xii{+})+\Gg(\Xikl)
-q_\Lambda\xii{0}^\Lambda \Gg'(\Xikl)+\hf\, q_\Lambda p^\Lambda \(\Gg'(\Xikl)\)^2 \).
\end{split}
\label{allT}
\ee
The corresponding functions associated to the pair of patches $\cU_-$ and $\cU_0$ can be obtained
replacing $F(\xii{+})$ by $\bF(\xii{-})$ and changing the sign of the function $\Gg$.
The explicit expression for $\Gg'(\Xikl)$ is given by
\be
\Gg'(\Xikl)= -\frac{\hnkl}{2\pi^2}\[
\int_0^{-\I\infty} \frac{\Xikl\,\de\Xi}{\Xikl^2-\Xi^2}\, \log\left(1-e^{-2\pi \I\, \Xi} \right)
+\frac{\pi}{12\I\Xikl}\]
\label{resultT}
\ee
and $\Xikl$ can be found in \eqref{resXig}.
Writing $\frac{2\,\Xikl}{\Xikl^2-\Xi^2}=\frac{1}{\Xikl-\Xi}+\frac{1}{\Xikl+\Xi}$,
it is clear that $\Gg'$ considered as a function of $\varpi$, as well as the initial function $\Gg$,
is a sum of two functions each having two cuts
going from $\varpi=0$ and $\varpi=\infty$ to two zeros of $\Xikl$, which are supposed
to lie outside $\cU_+\cup\cU_-$.

\EPSFIGURE{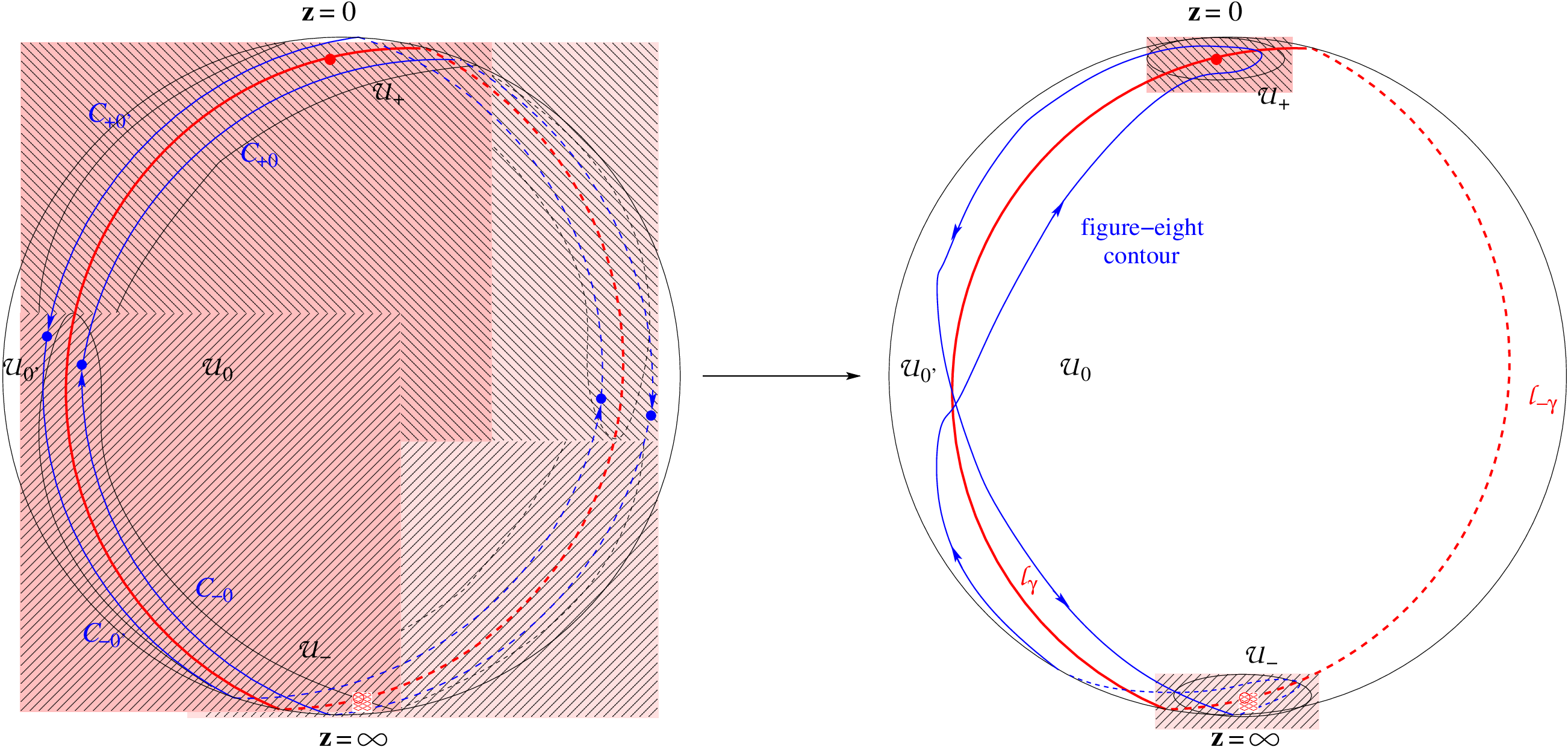,height=7.5cm}{The contours on $\CP$. The left picture shows
the covering of $\CP$ and the contours used in \eqref{txiqline_rer}.
On the right picture these contours are transformed into one figure-eight contour, which should be used
to integrate one of the two terms in $\Gg'$. The second term should be integrated along a similar
figure-eight contour going around $\ellg{-\gamma}$.
\label{fig_cont}}

Due to the presence of these cuts, the representation \eqref{txiqline}
for the twistor lines is not directly applicable because the contours around $\cU_\pm$
cannot be closed. To overcome this problem, it is convenient to start with the following equivalent
representation, exemplified here for $\txii{0}_\Lambda$,\footnote{Note a sign difference
with \eqref{txiqline}.}
\be
\txi_\Lambda^{[0]}= \frac{\I}{2}\, B_\Lambda
-\half  \sum_\pm \[\oint_{C_{\pm 0}} \frac{\de \varpi'}{2 \pi \I \varpi'} \,
\frac{\varpi' + \varpi}{\varpi' - \varpi}
\, \tilde{T}_\Lambda^{[\pm 0]}(\varpi')
+\oint_{C_{\pm 0'}} \frac{\de \varpi'}{2 \pi \I \varpi'} \,
\frac{\varpi' + \varpi}{\varpi' - \varpi}
\, \tilde{T}_\Lambda^{[\pm 0']}(\varpi')\] ,
\label{txiqline_rer}
\ee
where $\varpi\in \cU_0$ and the contours are shown on Fig. \ref{fig_cont} (left).
Then we have to consider separately terms
differing by the power of $\Gg$. The terms independent of $\Gg$, which are proportional to
the holomorphic prepotential, are all meromorphic,
having at most poles at $\varpi=0$ and $\varpi=\infty$.
Moreover, they coincide from two sides ($\cU_0$ and $\cU_{0'}$). Therefore, for these terms
the contours can be deformed just to two small circles around the north and south pole of $\CP$,
and the corresponding integrals are evaluated by residues.
All terms linear in $\Gg$, as noticed above, can be written as
a linear combination of functions with two cuts
and they have opposite signs in $T$'s with indices $\scriptstyle [\pm 0]$.
Such situation was analyzed in \cite{Alexandrov:2008ds} where it was demonstrated
that for each term the initial contours can be combined into one ``figure-eight" contour going around
$\varpi=0$ and $\varpi=\infty$, as shown on Fig. \ref{fig_cont} (right).
The integrals around such contours reduce then to
integrals along $\ellg{\pm\gamma}$ of the corresponding discontinuity.
This analysis is sufficient to evaluate $\xii{0}^\Lambda$ and $\txii{0}_\Lambda$.
It is easy to verify that the results indeed coincide with \eqref{xiqlineB2} and \eqref{txiqlineB2}.

The calculation of $\alpha^{[0]}$ is more complicated due to the presence of terms
quadratic in $\Gg$, which should be treated with great care.
Let us collect the contributions quadratic in instantons, which come from the last two terms
in $\tilde{T}^{[\pm 0]}_\alpha$. They are given by
\be
\frac{\I}{4}\, q_\Lambda p^\Lambda\( (\Gg')^2\mp \frac{\hnkl}{4\pi^2}\, \Ikl^{(1)}(\varpi)\Gg'\)
=\frac{\I}{4}\, q_\Lambda p^\Lambda\[\( \Gg'\mp \frac{\hnkl}{8\pi^2}\, \Ikl^{(1)}(\varpi)\)^2
-\frac{\hnkl^2}{64\pi^4} \(\Ikl^{(1)}(\varpi)\)^2\].
\label{quadrcuts}
\ee
The first term is regular in $\cU_\pm$, as was noticed in \eqref{an_combin}.
Thus, a non-trivial contribution to $\alpha^{[0]}$ originates from the second term only.
The latter has two cuts along $\ellg{\pm\gamma}$ and the sum of all integrals
reduce to an integral along these contours of the residue
of \eqref{quadrcuts}.\footnote{The additional contribution from the poles $\varpi=0,\infty$
of the integration measure in \eqref{txiqline} is canceled by a similar one from the first term.}
As a result, one finds the following contribution
\be
\begin{split}
& -\frac{\I\hnkl^2}{(4\pi)^4}\, q_\Lambda p^\Lambda \sum\limits_{s_1,s_2=\pm} s_1 s_2
\int_0^{\I s_1\infty}\cD^{(s_1)} [\varpi_1]
\int_0^{\I s_2\infty}\cD^{(s_2)} [\varpi_2]
\,\frac{2(\varpi_1+\varpi_2)\varpi}{(\varpi_1-\varpi)(\varpi_2-\varpi)}
\\
& =-\frac{\I\hnkl^2}{(4\pi)^4}\, q_\Lambda p^\Lambda \( \(\Ikl^{(1)}(\varpi)\)^2+16\Kkl^2  \)
,
\end{split}
\label{quadrinst}
\ee
where we denoted the measure
$\cD^{(s)} [\varpi] =\frac{\de\varpi}{\varpi}\,\log\left(1-e^{-2\pi \I s \Xikl(\varpi)}\right)$.
The other terms in $\tilde{T}^{[\pm 0]}_\alpha$ can be treated in the way discussed above.
The only subtlety is that reducing the integral along the figure-eight contour
for the third term in \eqref{allT} to the integral of its discontinuity, one should take
into account the contributions from the poles of the integration measure.
After lengthy calculations, one arrives at the total result, which
is conveniently formulated in terms of the combination $\tilde\alpha$ and
coincides with \eqref{txifqlineB2}.
In particular, the contribution quadratic in instantons \eqref{quadrinst}
is canceled in this combination and the linear instanton approximation is exact.
The only place where the quadratic terms appear is the definition of the field $\sigma$ in
\eqref{xzeta}.

\section{Verification of transition functions}
\label{ap_transpole}

The aim of this appendix is to check that the transition functions \eqref{transpole}
satisfy all necessary conditions, namely, that they define regular contact twistor lines
in the patch $\cU_+$ and they are consistent with the transition functions \eqref{transellg}.
But first we need to evaluate their derivatives. As usual, the situation is complicated by the fact
that they are written as functions of coordinates in the patch $\cU_+$, whereas
in the contact transformations they are considered as functions of $\xii{+}^\Lambda$ and $\txii{i}_\Lambda$.

Let us start by evaluating derivative of $\hHij{+i}$ w.r.t. $\Xipgi{m}$. After integrating
by parts in the second and forth terms, one finds
\bea
\label{derfirst}
{2\I}\,\frac{\p\hHij{+i}}{\p \Xipgi{m}}&=&
\frac{\hng{m}}{4\pi^2}  \(\int_0^{-\I\infty} \!\!\frac{\Dlog{m}}{\Xipgi{m}-\Xi}
+\frac{\pi \I}{12\Xipgi{m}}\)
\\
&+&
\frac{1}{(8\pi^2)^2}\sum_{j} \hng{j}q_{j,\Lambda}\,\frac{\p\derG{j}}{\p \Xipgi{m}}
\sum_{k} \hng{k}p_{k}^{\Lambda}\derG{k}
+\frac{1}{2(8\pi^2)^2}\sum_{j\ne k}\hng{j} \hng{k}\<\gamma_j,\gamma_k\>\derG{j}\,\frac{\p\cF_{k}}{\p \Xipgi{m}}
\nn \\
& -&
\frac{1}{8\pi^2}\sum_{j} \hng{j}
\int_0^{-\I\infty} \!\!\frac{\de \Xi}{\Xipgi{j}-\Xi}\[
\(\frac{\p\Sgi{j}(\Xi)}{\p \Xipgi{m}}+\delta_{j,m}\, \frac{\p\Sgi{j}(\Xi)}{\p \Xi} \)
 \right.
\nn \\
& - & \left.
\( \Sgi{j}(\Xi)-\Sgi{j}(\Xipgi{j})\)
\(\frac{\p}{\p \Xipgi{m}}+\delta_{j,m}\, \frac{\p}{\p \Xi} \)\]
\log\left(1-e^{-2\pi \I\( \Xi-\Sgi{j}(\Xi)\)} \right) .
\nn
\eea
Using equation \eqref{defSF} for $\Sgi{i}(\Xi)$, it is easy to obtain that the very last
term is equal to
\be
\textstyle
\frac{1}{32\pi^4}\sum\limits_{j\ne k}\hng{j} \hng{k}\<\gamma_j,\gamma_k\>
\int\limits_0^{-\I\infty} \!\!\frac{\Dlog{j}}{\Xipgi{j}-\Xi}
\int\limits_0^{-\I\infty}\!\! \frac{\Xi\,\de\Xi'}{\Xipgi{k}\Xi-\Xipgi{j}\Xi'}
\(\frac{\p}{\p \Xipgi{m}}+\delta_{j,m}\, \frac{\p}{\p \Xi'} \)
\log\left(1-e^{-2\pi \I\( \Xi'-\Sgi{k}(\Xi')\)}\right).
\label{lastterm}
\ee
On the other hand, the combination of the third and forth terms gives
\bea
\label{nextterm}
\textstyle
&& \frac{1}{32\pi^4}\sum\limits_{j\ne k}\hng{j} \hng{k}\<\gamma_j,\gamma_k\>\left\{
\frac{1}{4}\, \cF_j\,\frac{\p\cF_{k}}{\p \Xipgi{m}}
\right.
\\
&& \textstyle \left. +
\int\limits_0^{-\I\infty} \!\!\frac{\Dlog{j}}{\Xipgi{j}-\Xi}
\int\limits_0^{-\I\infty}\!\! \frac{\de\Xi'}{\Xipgi{k}\Xi-\Xipgi{j}\Xi'}
\[\frac{\delta_{j,m}\Xi'(\Xipgi{j}-\Xi)+\delta_{k,m}\Xi^2}{\Xipgi{k}\Xi-\Xipgi{j}\Xi'}
-\Xi\frac{\p}{\p \Xipgi{m}} \]
\log\left(1-e^{-2\pi \I\( \Xi'-\Sgi{k}(\Xi')\)}\right)\right\}.
\nn
\eea
Combining these two contribution and using eq. \eqref{regfun} for $\cF_j$, one arrives at the final result
\be
{2\I}\,\frac{\p\hHij{+i}}{\p \Xipgi{m}}=
\frac{1}{8\pi^2}\sum_k\hng{k}\(\delta_{k,m}+
\frac{1}{8\pi^2}\sum_{j} \hng{j}q_{j,\Lambda}\,\frac{\p\derG{j}}{\p \Xipgi{m}}\,p_{k}^{\Lambda}\)
 \derG{k}.
\label{finalder}
\ee
One immediately concludes that it leads to
\be
2\I\p_{\xii{+}^\Lambda}\hHij{+i}=
F_\Lambda(\xii{+}^\Lambda)+\frac{1}{8\pi^2}\sum_j\hng{j}q_{j,\Lambda}\derG{j},
\qquad
\p_{\txii{i}_\Lambda}\hHij{+i}=\frac{1}{8\pi^2}\sum_j\hng{j}p_{j}^{\Lambda}\derG{j}.
\label{derivH}
\ee

Now it is easy to verify that the contact transformation generated by $\hHij{+i}$
removes all singularities of the contact Darboux coordinates near the north pole.
First, the regularity of $\xii{+}^\Lambda,\txii{+}_\Lambda$ trivially follows from the
fact that their instanton contributions depend only on the combination
$\Igg{i}+\derG{j}$, which is regular around $\varpi=0$.
The analysis of $\alpi{+}$ is as usual a bit more complicated. It can be represented as
\be
\begin{split}
4\I \alpi{+}=&\,
\tilde \alpha^{[i]}-2\I\[ \txii{+}_\Lambda\xii{+}^\Lambda
+ \(2-\xii{+}^\Lambda\p_{\xii{+}^\Lambda}-\txii{+}_\Lambda\p_{\txii{i}_\Lambda}  \)\hHij{+i}
-\p_{\xii{+}^\Lambda}\hHij{+i}\p_{\txii{i}_\Lambda}\hHij{+i}\]
\\
=&\, \tilde \alpha^{[i]}-2\I \txii{+}_\Lambda\xii{+}^\Lambda
+\frac{\I}{4\pi^3}\sum_j\hng{j}
\int_0^{-\I\infty}\!\! \frac{\de\Xi}{\Xipgi{j}-\Xi}\, \Li_2\left(e^{-2\pi \I\(\Xi-\Sgi{j}\)} \right)
\\
&
+\frac{1}{8\pi^2}\sum_{j} \hng{j}\Xipgi{j} \derG{j}
-\frac{1}{4\pi^2}\sum_{j} \hng{j}
\int_0^{-\I\infty}\!\! \frac{\Dlog{j}}{\Xipgi{j}-\Xi}\, \Sgi{j}(\Xi)
.
\end{split}
\label{talp}
\ee
Note that in the forth term one can safely replace $\Xipgi{j}$ by $\Xigi{j}$.
Then the discontinuities of this term arising due to the branch cuts of $\Xigi{j}$ cancel those of the
last term in \eqref{talp}, whereas the discontinuities coming from the cuts of $\derG{j}$
cancel those of the two last terms in $\tilde \alpha^{[i]}$ \eqref{manyalpha}.
At the same time the third term in \eqref{talp} removes the singularities of the forth term in \eqref{manyalpha}.
Since the second term is regular, we conclude that $\alpi{+}$ does not have branch cut
singularities in $\cU_+$ except the term coming from the anomalous dimension.
Besides, it is easy to show that all simple poles at $\varpi=0$ also cancel
each other. Thus, the contact Darboux coordinate $\alpi{+}$ satisfies all regularity conditions.

Finally, let us check that the proposed form of $\hHij{+i}$ is consistent with
the transition functions through the BPS rays \eqref{transellg}. The latter
can be obtained from the composition law \eqref{comp}. The term linear in the transition
functions gives
\be
\frac{\I}{2}\( \Ggss{\gamma_i}-\hf \,q_{i,\Lambda}p_i^\Lambda\(\Ggss{\gamma_i}'\)^2
+\frac{\Ggss{\gamma_i}'}{16\pi^2}\sum_j \hng{j} \(q_{i,\Lambda}p_j^\Lambda+q_{j,\Lambda}p_i^\Lambda\)\derG{j}
+\frac{1}{2}\, \Ggss{\gamma_i}'\Sgi{i}(\Xipgi{i})\),
\ee
where $\derG{i}$ denotes the branch of the function \eqref{instXi} obtained by analytical continuation
from  the patch $\cU_i$ clockwise.
The quadratic term is evaluated using \eqref{derivH} and leads to
\be
-\frac{\I}{16\pi^2}\,q_{i,\Lambda}\Ggss{\gamma_i}'\sum_j\hng{j}p_{j}^{\Lambda}\derG{j}.
\ee
Altogether these contributions reproduce our initial starting point \eqref{transellg}.
This completes the verification that \eqref{transellg} and \eqref{transpole}
form a consistent set of transition functions.

\end{document}